\documentclass[oneside,a4paper,onecolumn,11pt]{article}

\usepackage{fullpage}
\usepackage[left=2cm,top=2cm,bottom=2cm,right=2cm,includehead,nomarginpar,headheight=16pt]{geometry}
\usepackage[ruled, linesnumbered, vlined, commentsnumbered]{algorithm2e}
\usepackage{graphicx,subfig} 
\usepackage{amsfonts,amssymb,amsmath,amsthm,amsopn,mathtools}	
\usepackage{booktabs,diagbox,colortbl,multirow,tabularx,threeparttable,hhline}
\usepackage[listings,skins,breakable]{tcolorbox}

\usepackage{fancyhdr,fancyheadings,nopageno,lastpage} 
\usepackage{enumerate}
\usepackage[shortlabels]{enumitem}
\usepackage{csquotes}
\usepackage{authblk}
\usepackage{footnote}
\usepackage[colorlinks=true,citecolor=reference,linkcolor=brightmaroon,backref=page]{hyperref}
\usepackage{prettyref}
\usepackage{cite}
\usepackage{setspace}
\usepackage{color}
\usepackage{titlesec}
\usepackage{soul}
\usepackage{url}
\usepackage{bm}
\usepackage{xcolor}  
\usepackage{csquotes}
\usepackage{academicons}
\usepackage{fontawesome5}
\usepackage{pifont,ifsym,marvosym,manfnt} 

\usepackage{tikz}
\usepackage{pgfplots}
\usetikzlibrary{positioning,shapes,shadows,arrows,calc}
\tikzstyle{component}=[rectangle, draw=black, rounded corners, fill=blue!40, drop shadow, text centered, anchor=north, text=white, minimum height=1cm]
\tikzstyle{arrow}=[->, thick]

\pgfplotsset{compat=1.12}
\usetikzlibrary{intersections}
\usetikzlibrary{pgfplots.statistics}
\usepgfplotslibrary{fillbetween}

\fancypagestyle{plain}{%
    \fancyfoot[C]{}
    \fancyfoot[R]{\faLeanpub\ \, \thepage\ / \pageref{LastPage}}
    \fancyfoot[L]{\faBraille\ \textsc{COLALab Report} \faSlackHash\ \footnotesize\textifsym{2024004}}
}

\hypersetup{
    colorlinks=true,
    linkcolor=ultramarine,
    filecolor=magenta,
    urlcolor=ultramarine,
    pdftitle={Overleaf Example},
    pdfpagemode=FullScreen,
}

\definecolor{red(munsell)}{rgb}{0.95, 0.0, 0.24}
\definecolor{navyblue}{RGB}{0, 0, 128}
\definecolor{myblue}{RGB}{34,31,217}
\definecolor{mycyan}{gray}{.7}
\definecolor{Gray}{gray}{0.9}
\definecolor{usccardinal}{rgb}{0.6, 0.0, 0.0}
\definecolor{ultramarine}{RGB}{0,32,96}
\definecolor{amber}{rgb}{1.0, 0.49, 0.0}
\definecolor{greenstar}{RGB}{83, 182, 85}
\definecolor{purplestar}{RGB}{96, 55, 148}
\definecolor{reference}{RGB}{4, 20, 110}
\definecolor{linenum}{RGB}{21, 127, 127}
\definecolor{amaranth}{rgb}{0.9, 0.17, 0.31}
\definecolor{brightmaroon}{rgb}{0.76, 0.13, 0.28}
\definecolor{lightpurple}{RGB}{242, 239, 246}
\definecolor{f1green}{RGB}{97,182,92}
\definecolor{f1blue}{RGB}{59,92,152}
\definecolor{f3purple}{RGB}{177,142,191}
\definecolor{f3blue}{RGB}{99,128,177}

\newsavebox\newcaptionbox\newdimen\newcaptionboxwid
\titlespacing*{\paragraph}{0pt}{0.75ex plus 0.75ex minus 0.2ex}{1.25ex plus 0.2ex}

\long\def\@makecaption#1#2{
 \vskip 10pt
        \baselineskip 11pt
        \setbox\@tempboxa\hbox{#1. #2}
        \ifdim \wd\@tempboxa >\hsize
        \sbox{\newcaptionbox}{\small\sl #1.~}
        \newcaptionboxwid=\wd\newcaptionbox
        \usebox\newcaptionbox {\footnotesize #2}
        \else
          \centerline{{\small\sl #1.} {\small #2}}
        \fi}

\newtcolorbox{quotebox}{
    colback=lightpurple,
    colframe=black!75,
    boxrule=0pt,
    top=5pt,
    bottom=5pt,
    left=8pt,
    right=8pt,
    arc=8pt,
    boxsep=0pt,
    toptitle=2pt,
    bottomtitle=2pt,
    fonttitle=\bfseries,
}

\SetCommentSty{mycommfont}

\DeclareMathOperator*{\argmax}{argmax}
\DeclareMathOperator*{\argmin}{argmin}

\newcommand{\pref}{\prettyref}

\newrefformat{fig}{Fig.~\ref{#1}}
\newrefformat{tab}{Table~\ref{#1}}
\newrefformat{sec}{Section~\ref{#1}}
\newrefformat{alg}{Algorithm~\ref{#1}}
\newrefformat{property}{Property~\ref{#1}}
\newrefformat{theorem}{Theorem~\ref{#1}}
\newrefformat{definition}{Definition~\ref{#1}}
\newrefformat{corollary}{Corollary~\ref{#1}}
\newrefformat{lemma}{Lemma~\ref{#1}}
\newrefformat{conj}{Conjecture~\ref{#1}}
\newrefformat{def}{Definition~\ref{#1}}
\newrefformat{eq}{equation~(\ref{#1})}
\newrefformat{app}{Appendix~\ref{#1}}

\usepackage{lscape}

\newenvironment{code-example}
{
\vspace{0.15cm}
\noindent\begin{minipage}{\linewidth}
\begin{center}
\arrayrulecolor{black}
\color{black}
\begin{tabular}{|p{0.95\linewidth}|}
\hline%
\rowcolor{pink!20}%
}
{
\\\hline
\end{tabular}
\end{center}
\end{minipage}
\vspace{-0.2cm}
}

\begin{document}

\title{\vspace{-1ex}\LARGE\textbf{Unlocking the Secrets of Software Configuration Landscapes---Ruggedness, Accessibility, Escapability, and Transferability}~\footnote{This manuscript is currently submitted for possible publication. Reviewers can feel free to use this in peer review.}}

\author[1]{\normalsize Mingyu Huang$^\dagger$}
\author[1]{\normalsize Peili Mao\footnote{Mingyu Huang and Peili Mao contributed equally to this work.}}
\author[2]{\normalsize Ke Li}
\affil[1]{\normalsize School of Computer Science and Engineering, UESTC, Chengdu, PR China}
\affil[2]{\normalsize Department of Computer Science, University of Exeter, EX4 4QF, Exeter, UK}
\affil[\Faxmachine\ ]{\normalsize \texttt{k.li@exeter.ac.uk}}

\date{}
\maketitle

\vspace{-3ex}
{\normalsize\textbf{Abstract: } }Modern software systems are often highly configurable to tailor varied requirements from diverse stakeholders. Understanding the mapping between configurations and the desired performance attributes plays a fundamental role in advancing the controllability and tuning of the underlying system, yet has long been a dark hole of knowledge due to their black-box nature and the enormous combinatorial configuration space. In this paper, using $86$M evaluated configurations from three real-world systems on $32$ running workloads, we conducted one of its kind fitness landscape analysis (FLA) for configurable software systems. With comprehensive FLA methods, we for the first time show that: $i)$ the software configuration landscapes are fairly rugged, with numerous scattered local optima; $ii)$ nevertheless, the top local optima are highly accessible, featuring significantly larger basins of attraction; $iii)$ most inferior local optima are escapable with simple perturbations; $iv)$ landscapes of the same system with different workloads share structural similarities, which can be exploited to expedite heuristic search. Our results also provide valuable insights on the design of tailored meta-heuristics for configuration tuning; our FLA framework along with the collected data, build solid foundation for future research in this direction.



\section{Introduction}
\label{sec:introduction}

Modern software systems have become increasingly sophisticated and highly configurable. For example, the \textsc{Linux} kernel has $15,000+$ options, most of which can take three values. Without considering constraints, this can yield $\approx 3^{15,000}$ configurations---far more than the number of atoms in the universe ($\approx 10^{80}$). On one hand, by tuning various configuration options, stakeholders are empowered with the flexibility to tailor a system according to their requirements on different non-functional performance attributes (e.g., execution time, throughput)~\cite{XuJFZPT15,SiegmundGAK15,VelezJSSAK20}; yet on the other hand, stakeholders are usually overwhelmed by the complexity of the system under tune and ending up with innate settings or relying on expert consultations. A thorough understanding of the intricate relationship between configurations and performance is essential for effective control of the system under tune, facilitating various downstream tasks such as analysis~\cite{VelezJSSAK20}, debugging~\cite{HanYL18}, adaptation~\cite{ChenLBY18}, optimization~\cite{ZhuLGBMLSY17,Nair0MSA20}, and autotuning~\cite{AshouriKCPS19}. However, this is far from trivial in practice due to the black-box nature of software systems.

The fitness landscape metaphor, pioneered by Wright in 1932~\cite{Wright32}, is a fundamental concept in evolutionary biology~\cite{VaishnavBMYFATLCR22}, and has been adapted for analyzing input-output responses of black-box systems across various disciplines, e.g., computer science \cite{Malan21}, social science \cite{Levinthal91}, chemistry \cite{Doye02}, and engineering \cite{VassilevMF99}. It can be envisioned as a (hyper-)surface as formed by \textit{fitness} (i.e., performance attribute in our context) values across a high-dimensional configuration space. Each spatial location within this landscape represents a configuration, with its \textit{height} indicating the fitness and optimal configurations sitting on the peaks.

Since a configuration optimization process is akin to navigating uphill towards these peaks, the topography of fitness landscapes (e.g., the location of fitness peaks and their interconnectivity) plays a pivotal role for understanding both system properties and algorithmic behavior. While a plethora of fitness landscape analysis (FLA) \cite{Malan21} methods have been developed over the past decades for exploring diverse landscape properties, little has been known about the topography of software configuration landscapes. This gap can be partially attributed to the enormous configuration space and the expensive cost of performance evaluation, which hinder the construction of high-fidelity landscapes. Additionally, the variety of use cases, a.k.a., workloads, in real-world engineering poses further challenges in drawing general conclusions.

In this paper, by mapping $32$ large configuration landscapes of three representative real-world systems, and by leveraging diverse FLA methods, we substantially advance the understanding of the intricacies of black-box configurable software systems from the following four fundamental aspects.

\paragraph{Ruggedness.} A fitness landscape can be either unimodal or multimodal, a.k.a., \textit{rugged}, where ruggedness can pose significant challenges for optimizers in locating the global optimum~\cite{ZhuLGBMLSY17}. Configuration landscapes of various software systems are reported to be highly rugged, characterized by local optima~\cite{JamshidiC16}. However, this phenomenon is typically observed during the optimization process, which explores only a small fraction of the entire configuration space. In addition, important properties of local optima, such as their spatial distribution and fitness, have remained largely underexpored. This then leads us to our first research question (RQ): \textbf{RQ1:} \textit{\lq\ul{How rugged are the configuration landscapes, and how are local optima distributed across these landscapes?}\rq}. By applying several classic FLA metrics, we demonstrate that local optima are widespread and randomly dispersed in configuration landscapes, contributing to a highly rugged topography (\pref{sec:rq1}).

\paragraph{Accessibility of global peaks.} In such rugged landscapes, featured in tens of thousands of local optima, locating the most prominent peaks might seem as hard as finding a needle in a haystack. While it has been widely known that the probability of encountering a local optimum during optimization depends on its size of basin of attraction, the following fundamental question remains unclear: \textbf{RQ2}: \textit{\lq\ul{whether global peaks in configuration landscapes have tiny, needle-like basins, or are their basins larger than others?}\rq}. Our analyses on over $1$M local optima and large-scale simulation, reveals that the ruggedness does not preclude the top local optima to be highly accessible: they typically feature significantly larger basins, which allow them to be reached with a higher probability (\pref{sec:rq2}). 

\paragraph{Escaping inferior peaks.} Despite this, it is not uncommon in practice that optimizers can sometimes get trapped by less-fit local optima. Then, a natural follow-up question is \textit{\lq\ul{whether we can escape such undesired local optima, and thus transition to superior ones}\rq}, which constitutes our \textbf{RQ3}. We investigated this by analyzing the local optima network (LON) \cite{OchoaTVD08} of each landscape, which is a well-established FLA tool for modeling the interconnectivity pattern between local optima. We observe that inferior local optima can be easily escaped by applying some simple perturbations, and possibly transition to the top peaks (\pref{sec:rq3}).

\paragraph{Landscape similarity.} It is recently shown that landscapes of the same family of black-box optimization problems may share inherent \textit{structural similarities}~\cite{HuangL23a}, which provides fundamental support for the effectiveness of transfer learning methods~\cite{PanY10}. Here, we hypothesize that this is also true for configurable software systems, and thus our \textbf{RQ4:} \textit{\lq\ul{Do configuration landscapes of the same system with different workloads share any structural similarities?}\rq} By quantifying the consistency of configuration ranks across distinct workloads, and the similarity of learned embeddings for LONs, we find inspiring evidence to support our hypothesis. We further show that such similarities can be exploited to expedite the optimization process (\pref{sec:rq4}).

To the best of our knowledge, this work represents the first attempt to conduct exploratory analysis on the landscapes of configurable software systems. Towards this end, 
$\blacktriangleright$ we develop a highly-scalable FLA framework that can manipulate landscapes with millions of configurations (\pref{sec:methods}). 
$\blacktriangleright$ We collect extensive performance data from $3$ prevalent software systems, with $32$ workloads, and over $86$M configurations in toal (\pref{sec:methods}). 
$\blacktriangleright$ We for the first time provide insights into four fundamental aspects of configuration landscapes.


\section{Background}
\label{sec:background}



\subsection{Software Configuration Problem}

A configurable software system often comes with a set of configuration options $\{c_1, \dots, c_n\}$, each of which can take either categorical (or boolean) or integer values. The whole configuration space, $\mathcal{C} = C_1 \times \dots \times C_n$, is the Cartesian product of the domains of all $n$ options of interest. Let $f: \mathcal{C} \to \mathbb{R}$ be a fitness function that maps from a configuration vector $\bm{c} = (c_1,\cdots,c_n) \in \mathcal{C}$ to a performance attribute such as execution time. The goal of software configuration is to find an optimum configuration $\bm{c}^* = \underset{\bm{c}\in\mathcal{C}}{\argmax/\argmin}~f(\bm{c})$. Due to the lack of analytical form of $f$, this is a black-box optimization problem that is known to be $\mathcal{NP}$-hard~\cite{Weise09}.

\subsection{Fitness Landscape}
\label{sec:FLA}

\paragraph{Fitness landscape.}
\label{sec:landscape}

A \textit{fitness landscape} can be defined as a triplet $(\mathcal{C}, \mathcal{N}, f)$, where $\mathcal{C}$ is the search space; $\mathcal{N}$ indicates a neighborhood structure that specifies the neighbors of each configuration; and $f: \mathcal{C} \rightarrow \mathbb{R}$ is the fitness function. 

\paragraph{Local optimum.} A configuration $\bm{c}$ is said to be a \textit{local optimum} (denoted as $\bm{c}^\ell$) iff $f(\bm{c}^\ell)$ is better than ${f(\bm{c})}$, $\forall\bm{c}\in{\mathcal{N}(\bm{c}^\ell)}$, where $\mathcal{N}(\bm{c}^\ell)$ is the neighborhood of $\bm{c}^\ell$.

\paragraph{Basin of attraction.} The \textit{basin of attraction} of a local optimum $\bm{c}^\ell$, denoted as  $\mathcal{B}(\bm{c}^\ell)$, is the set of all configurations from which local search converges to it, i.e., $\mathcal{B}(\bm{c}^\ell) =\{\bm{c}\in{\mathcal{C}}\mid\mathrm{LocalSearch}(\bm{c})\rightarrow\bm{c}^\ell\}$. We define the \textit{size} of a basin to be its cardinality $|\mathcal{B}|$, and its average \textit{radius} to be the expected number of local search steps to arrive at the corresponding local optimum. Note that depending on the local search strategy, e.g., \textit{best-} or \textit{first-}improvement~\cite{WhitleyHH13}, the exact basin size and radius can vary.

\paragraph{Local optima network.} LON is rooted in the study of energy landscapes in chemical physics~\cite{Stillinger95} and is a compact graph representation of fitness landscape (denoted as $\mathcal{G}=(\mathcal{V},\mathcal{E})$). It compresses all configurations in the basin of a local optimum $\bm{c}^\ell_i$ into a single vertex $v_i \in \mathcal{V}$ in LON. Each edge $e^{i,j}\in\mathcal{E}$ indicates potential transitions between the local optimum basins, i.e., whether a configuration in $\mathcal{B}(\bm{c}^\ell_i)$ can \textit{escape} to another basin $\mathcal{B}(\bm{c}^{\smash{\ell}}_j)$ by applying some perturbations. Since LON is able to capture various characteristics regarding LOs and their connectivity patterns, it has become one of the most popular methods for FLA~\cite{Malan21}.


\section{Configuration Landscape Construction}
\label{sec:methods}

This section first introduces our considered software systems in this study, along with the selected options and performance metrics (\pref{tab:setup}). We then delineate our data collection and landscape construction procedure. Due to the stringent page limit, we left more details in \textsc{Appendix} $1$ and \textsc{Appendix} $2$. 

\subsection{Software Systems and Data Collection}

\paragraph{Systems under tune.} To ensure practicality of our findings, in this paper we investigated three widely used configurable software systems~\cite{SiegmundGAK15}, including the $\blacktriangleright$ \textsc{LLVM} compiler, the $\blacktriangleright$ \textsc{Apache} web server, and the $\blacktriangleright$ \textsc{SQLite} database. Additionally, to account for varied real-world engineering use cases, we considered different workloads (\texttt{W}) for each system, which results in a total of $32$ different scenarios (see \textsc{Appendix} $1$).

\paragraph{Feature selection.} While each of these systems could have numerous configurable options, not all of them significantly impact performance. Thus, we employed a two-stage feature selection process to identify a more relevant subset of options: $\blacktriangleright$ Initially, we conducted a pre-selection based on each system's official documentation, including only those options explicitly stated to affect performance. $\blacktriangleright$ Then, we performed an ablation analysis~\cite{BiedenkappLEHFH17} to further filter out options without a statistically significant impact on performance. Following this approach, we finally identified $20$ options for \textsc{LLVM}, $18$ for \textsc{Apache}, and $16$ for \textsc{SQLite}.

\paragraph{Performance metrics.} In this study, we considered the performance metrics as in the official benchmarking toolkit of each system (see links below). Specifically, we analyze the $\blacktriangleright$ execution time of compiled projects for \textsc{LLVM}\href{https://polly.llvm.org/documentation/passes.html}{$^{\text{\tiny \textcolor{brightmaroon}{\smash{\faLink}}}}$}, the $\blacktriangleright$ number of requests handled per second for \textsc{Apache}\href{https://httpd.apache.org/docs/2.4/programs/ab.html}{$^{\text{\tiny \textcolor{brightmaroon}{\smash{\faLink}}}}$}, and the $\blacktriangleright$ number of written items per second for \textsc{SQLite}\href{https://www.oracle.com/docs/tech/berkeley-db-perf.pdf}{$^{\text{\tiny \textcolor{brightmaroon}{\smash{\faLink}}}}$}.

\paragraph{Data collection.} To collect benchmark data that is compatible with FLA, we exhaustively explored all combinations of categorical options and used a grid search for integer options\footnote{All relevant data can be found in \url{http://tinyurl.com/4vukrtct}.}. This approach resulted in $1.7$M configurations for each workload of \textsc{LLVM}, $2.9$M for \textsc{Apache}, and $5.0$M for \textsc{SQLite}. For each configuration, we conducted five runs to evaluate the targeted performance metrics and recorded the median value as the final performance indicator. It is important to note that configuration options not considered in this study were set to their recommended values during this data collection phase.

\begin{table}[t!]
    \caption{Meta-information of our experiments.}
    \centering
    \small
    \begin{tabular}{||lccc|c||}
        \hline
        System & Workloads & Options & Configs. & Total Eva. \\
        \hline
        \hline
        \textsc{LLVM}     & $12$ & $20$ & $1.05$M & $12.58$M \\
        \textsc{Apache}   & $10$ & $18$ & $1.77$M & $17.69$M \\
        \textsc{SQLite}   & $10$ & $16$ & $5.67$M & $56.65$M \\
        \hline
        \hline
        \textsc{Total}:            & $\bm{32}$ &  &  & $\bm{86.92\textbf{M}}$ \\
        \hline
    \end{tabular}
    \label{tab:setup}
\end{table}

\subsection{Construction Method}

\paragraph{Distance measures and neighborhood structure.} Central to our configuration landscape is a proper notion of \textit{distance} $d(\bm{c}_i,\bm{c}_j)$ between two configurations where $\bm{c}_i,\bm{c}_j\in\mathcal{C}$. For categorical (boolean) variables, we employ the Hamming distance as the measure, while for integer variables, the distance is defined as the number of steps between values on their respective grids. Consequently, $d(\bm{c}_i,\bm{c}_j)$ can be calculated as the sum of the distances for each corresponding pair of values. Given this, the neighborhood $\mathcal{N}(\bm{c})$ of a configuration $\bm{c}$ is then defined as the set of all configurations differing from $\bm{c}$ by exactly one distance, i.e., $\mathcal{N}(\bm{c})=\{\bm{c}^\prime\mid d(\bm{c^\prime},\bm{c})=1\}$.

\paragraph{Configuration landscape as a graph.} Given the inherent neighborhood structure in fitness landscapes, representing the data as a \textit{graph} is a natural choice. In this graph model, each configuration $\bm{c}\in\mathcal{C}$ is represented as a vertex, with the performance metric $f(\bm{c})$ asigned as the node attribute. Neighboring configurations $\bm{c}^\prime\in\mathcal{N}(\bm{c})$, are connected to $\bm{c}$ via \textit{directed edge}. The direction of each edge is determined by the relative values of $f(\bm{c})$ and $f(\bm{c}^\prime)$, always pointing towards the \textit{fitter} configuration. This graph-based approach to modeling the configuration landscape then enables the identification of local optima and their basins, as well as the implementation of FLA methods, using straightforward graph traversal techniques (see \textsc{Appendix} $2$). This allows our FLA framework to efficiently explore probably the largest landscapes ever in existing literature, comprising millions of configurations.

\paragraph{LON construction.} The essence of LON construction lies in defining the transitions between local optima, following the established routine in~\cite{OchoaV16} (\pref{alg:lon}). Specifically, for each local optimum $\bm{c}^\ell \in \mathcal{C}$, we apply a $k$-kick ($k\geq 2$) perturbation to it, resulting in a new configuration $\bm{c}^{\prime\prime}$ where $d(\bm{c}^\ell,\bm{c}^{\prime\prime})=k$. A local search initiated from $\bm{c}^{\prime\prime}$ leads to a new local optimum $\bm{c}^\ell_{\mathrm{new}}$. If $\bm{c}^\ell_{\mathrm{new}}$ is identical to $\bm{c}^\ell$, the process fails to \textit{escape} from $\bm{c}^\ell$. Otherwise, an edge is drawn from $\bm{c}^\ell$ to $\bm{c}^\ell_{\mathrm{new}}$. In this study, we examine all $2$-kick perturbations for each local optimum and record transition frequencies as edge weights. Additionally, following~\cite{ThomsonDO17}, we retain only \textit{improving edges}, i.e., those leading to a local optimum with better fitness, which results in a so-called \textit{monotonic} LON.

\begin{algorithm}[t!]
    \small
    \caption{Constructing local optima network}
    \label{alg:lon}

    \KwIn{The set of local optima $\mathcal{V}$; Configuration space $\mathcal{C}$; A neighborhood function $\mathcal{N}_d(\mathbf{c})$}
    \KwOut{$\mathcal{G} = (\mathcal{V}, \mathcal{E}, \mathcal{W})$}

    \ForEach{$\mathbf{c}^\ell \in \mathcal{V}$}{
        \ForEach{$\mathbf{c}^{\prime\prime} \in \mathcal{N}_2(\mathbf{c}^\ell)$}{
            $\mathbf{c}^\ell_{\text{new}} \gets \textsc{LocalSearch}(\mathbf{c}^{\prime\prime})$\;
            \If{$f(\mathbf{c}^\ell_{\text{new}}) < f(\mathbf{c}^\ell)$ 
                \textbf{and} 
                $\mathbf{c}^\ell_{\text{new}} \neq \mathbf{c}^\ell$}{
                $\mathcal{E} \gets \mathcal{E} \cup \{(\mathbf{c}^\ell, \mathbf{c}^\ell_{\text{new}})\}$; $\mathcal{W}[(\mathbf{c}^\ell, \mathbf{c}^\ell_{\text{new}})] \gets 1$;
            }
            \Else{
                $\mathcal{W}[(\mathbf{c}^\ell, \mathbf{c}^\ell_{\text{new}})] \gets \mathcal{W}[(\mathbf{c}^\ell, \mathbf{c}^\ell_{\text{new}})] + 1$;
            }
        }
    }
\end{algorithm}



\section{Configuration Landscape Analysis}
\label{sec:results}

This section addresses the RQs outlined in~\pref{sec:introduction} with a series of dedicated FLA methods. For sake of brevity, some per-instance discussions will focus on the \textsc{LLVM}-\texttt{W1}. Note that this is not indicative of bias: as detailed in \textsc{Appendix} $3$, our findings are representative across all $32$ scenarios.

\subsection{Configuration Landscapes are Highly Rugged}
\label{sec:rq1}

\begin{figure}[t!]
    \centering
    \includegraphics[width=.5\linewidth]{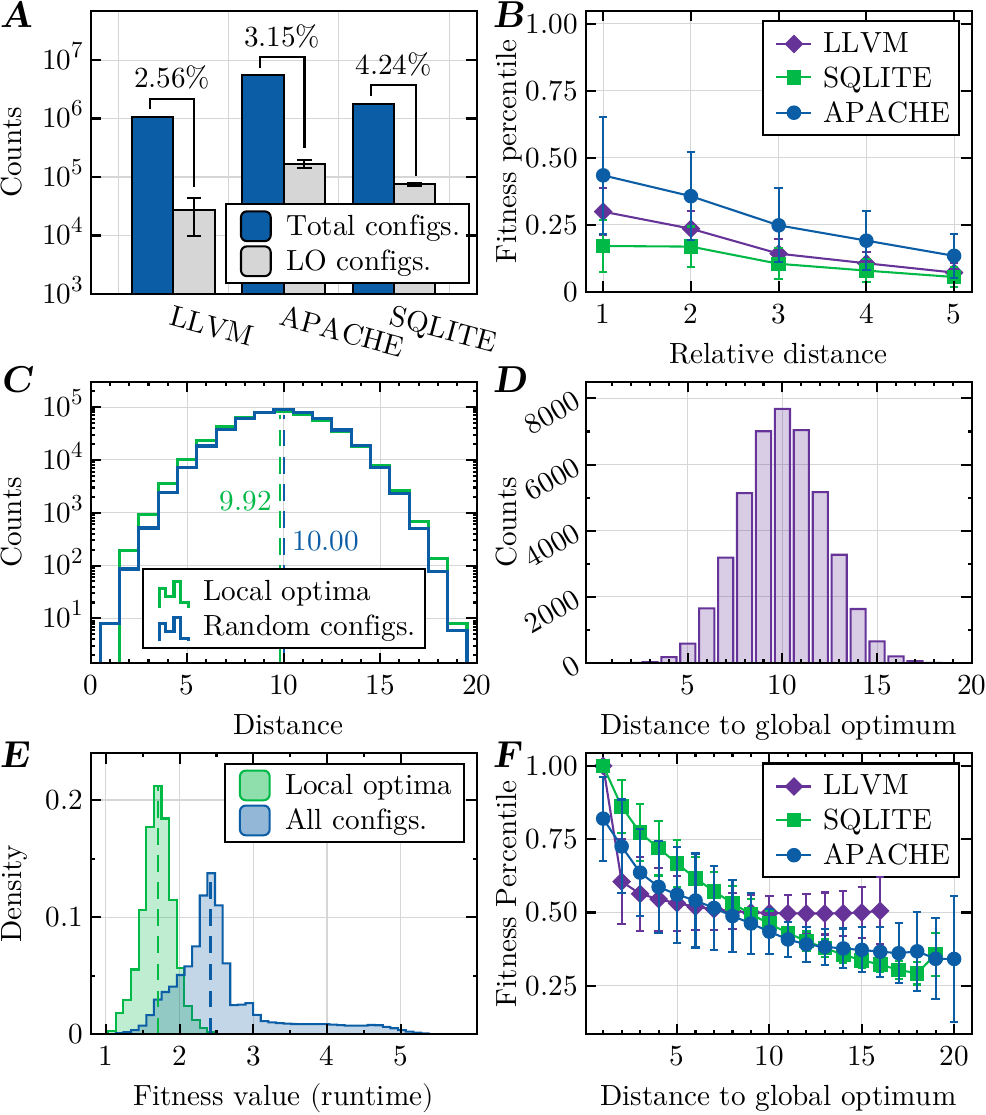}
    \caption{\textbf{(A)} Number of total configurations and local optima in each system, aggregated across workloads. \textbf{(B)} Autocorrelation calculated under different distances, aggregated across workloads. \textbf{(C)} Distribution of pairwise distance between local optima and $10^4$ randomly sampled configurations for \textsc{LLVM}-\texttt{W1}, where dashed lines are means. \textbf{(D)} Distribution of the distance of each local optimum to the global optimum for \textsc{LLVM}-\texttt{W1}. \textbf{(E)} Distribution of fitness values of all configurations local optima for \textsc{LLVM}-\texttt{W1}. \textbf{(F)} Average local optimum fitness (in percentile, lower is better) versus the corresponding distance to the global optimum for each system.}
    \label{fig:rq1}
\end{figure}

In order to investigate \textbf{RQ1}, we evaluated the number of local optima in each landscape and the autocorrelation metric. We also analyzed the spatial distribution of local optima as well as their fitness values, and thereby offer a comprehensive assessment of the landscape's ruggedness. 

\paragraph{Number of local optima.} We first quantified the number of local optima in each constructed landscape, which serves as a coarse-grained indicator of landscape ruggedness. We found that all investigated landscapes exhibit a significant number of local optima, ranging from $10^4$ to $10^5$, which take on average $2.56\%$ to $4.24\%$ of the total configurations (see~\pref{fig:rq1}A). This indicates a high degree of ruggedness in the software configuration landscapes that is comparable to the maximally rugged Kauffman's $NK$-landscapes~\cite{Kauffman93}. 

\paragraph{Autocorrelation.} To further validate the observed ruggedness, we calculated the \textit{autocorrelation} metric~\cite{Weinberger90} for each landscape, which is a widely used method for rigorously assessing landscape ruggedness. This metric evaluates the extent to which configurations in close proximity (in terms of distance) tend to exhibit similar fitness values. It is defined by the autocorrelation of fitness values across configurations visited during random walks through the landscape. The corresponding results show a high concordance with the large number of local optima (see~\pref{fig:rq1}B), indicating that even in the most local regions (i.e., $d=1$), the fitness of configurations is only weakly correlated. This correlation diminishes further as the distance between configurations increases.

\paragraph{Distribution of local optima.} In landscapes with numerous local optima, understanding their distribution is crucial. In particular, we seek to ascertain whether they are closely clustered or widely dispersed. To investigate this, we analyzed the distribution of pairwise distances between local optima and compared it with that of $10$K randomly sampled configurations from each landscape. Surprisingly, across all scenarios, the two distributions were remarkably similar (two-sided Kolmogorov-Smirnov test $D<0.147$, $p>0.954$; see~\pref{fig:rq1}C). This suggests that local optima are almost uniformly distributed across the landscapes. While this analysis considers distances between all pairs of local optima, we also examined particular regions of interest. For example, \pref{fig:rq1}D depicts the distribution of distances between each local optimum and the global optimum in \textsc{LLVM}-\texttt{W1}, which reveals that most local optima are located far from the global peak, following a normal distribution with a median distance of $10$---the same as the radius of the \textsc{LLVM} landscape. 

\paragraph{Fitness distance correlation.} In addition to their spatial locations, the fitness distribution of local optima is also of significant importance in rugged landscapes. For instance, landscapes where the fitness of a local optimum negatively correlates with its distance to the global optimum can be \lq deceptive\rq~\cite{Mitchell98}, potentially misleading the optimizer. In our cases, we found that most local optima in configuration landscapes are fitter than random configurations, but only a few approach the fitness of the optimum (see~\pref{fig:rq1}E). In general, sub-optimal local optima are located in the vicinity of the global optimum, and their fitness generally decreases with increasing distance (see~\pref{fig:rq1}F). However, such correlation, a.k.a., fitness distance correlation~\cite{JonesF95}, is weak for most cases ($|r|<0.2$). Thus, it is not unusual in these landscapes to find moves that increase fitness but paradoxically lead further away from the true optimum.

\begin{quotebox}
    \noindent
    \faLightbulb \, \textbf{Response to RQ1:} \textit{Software configuration landscapes are highly rugged, where local optima are prevalent and nearly randomly dispersed. Most of them are far away from the global optimum, and as they approach the optimum, their fitness tend to increase.}
\end{quotebox}

\subsection{Global Peaks are Highly Accessible}
\label{sec:rq2}

\begin{figure}[t!]
    \centering
    \includegraphics[width=.55\linewidth]{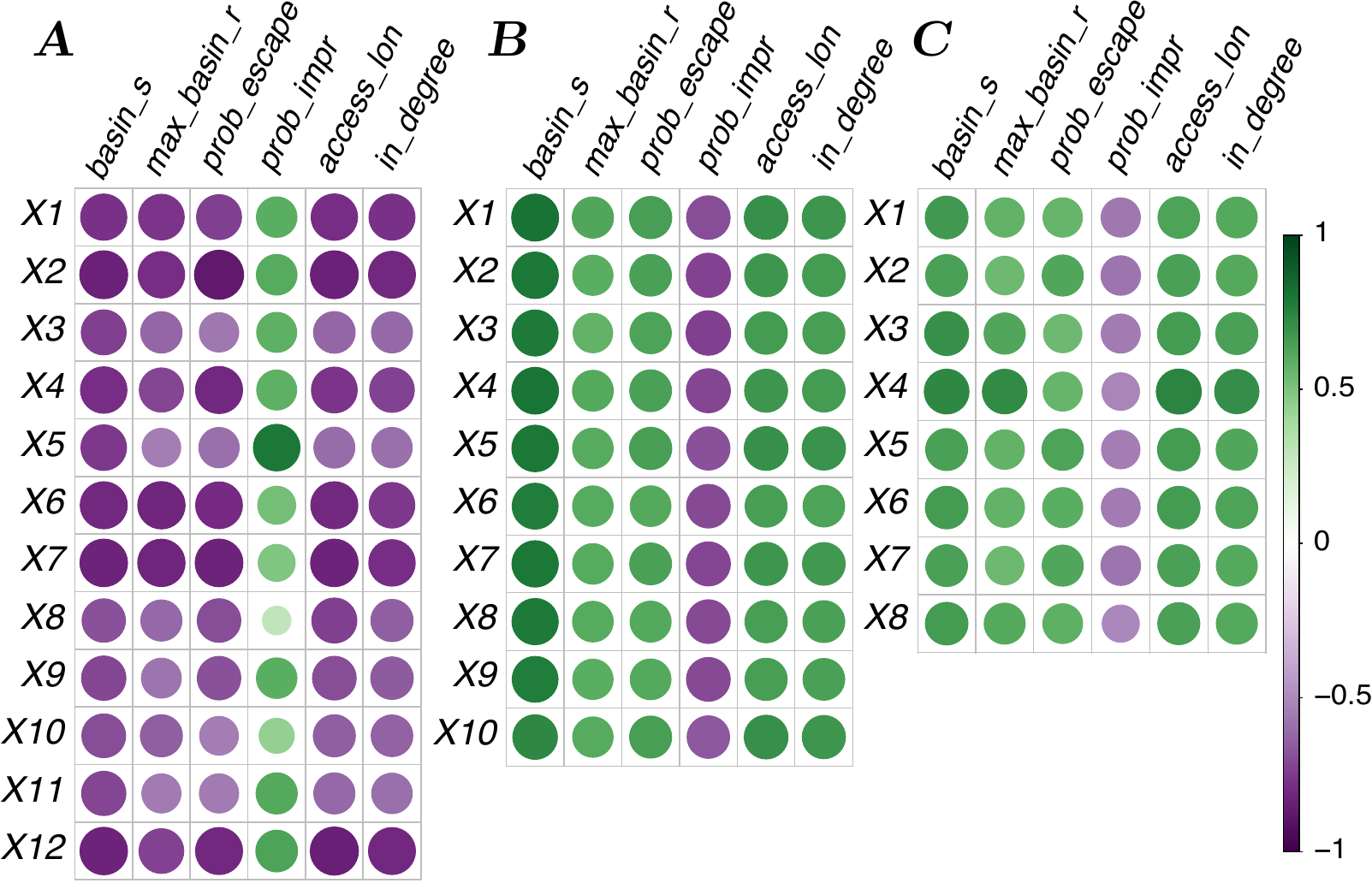}
    \caption{Spearman correlation between local optima fitness and their properties for \textbf{(A)} \textsc{LLVM}, \textbf{(B)} \textsc{SQLite}, and \textbf{(C)} \textsc{Apache}.}
    \label{fig:rq2_corr}
\end{figure}

\begin{figure}[t!]
    \centering
    \includegraphics[width=.5\linewidth]{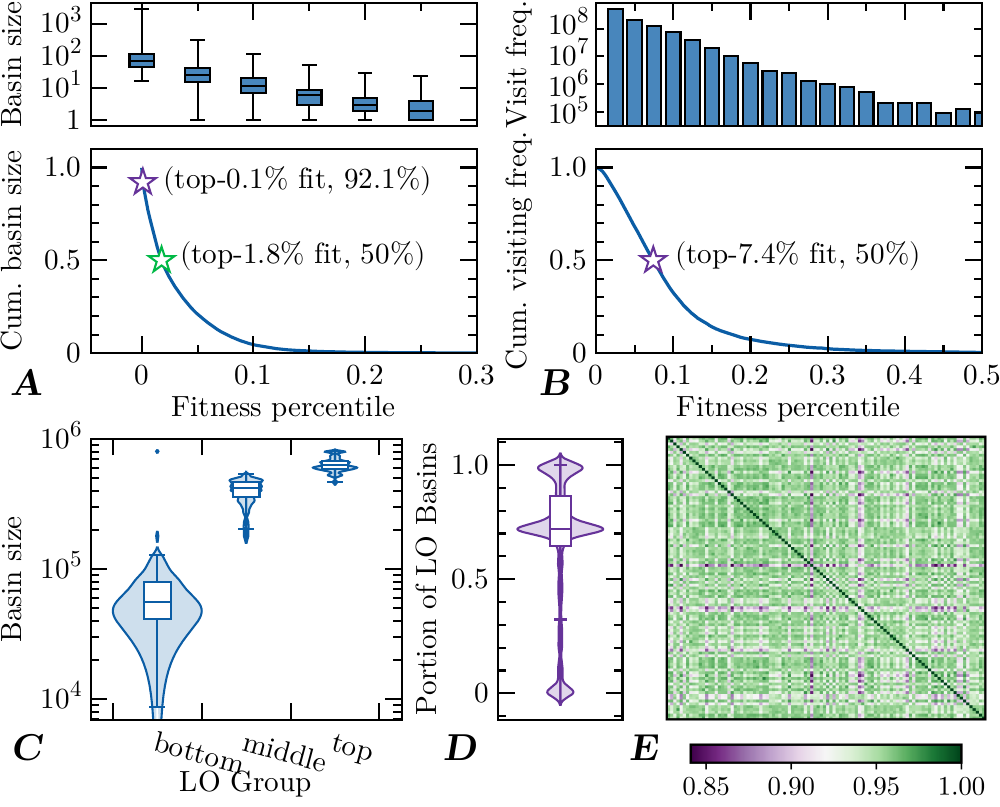}
    \caption{Example results on \textsc{LLVM}-\texttt{W1}: \textbf{(A)} boxplots and cumulative distribution of the size of best-improvement basins versus fitness percentile (the lower the better). \textbf{(B)} bars and cumulative distribution of frequency of visits for local optima at different fitness percentiles, during $10^9$ first-improvement local search runs. The long tails of both curves have been truncated. \textbf{(C)} distribution of basin size for local optima with fitness in the top, middle and bottom $0.1\%$ percentile based on first-improvement local search. \textbf{(D)} the distribution of accessibility to local optima basins of $10^3$ random configurations, where $1.0$ implies a configuration is shared by all basins. \textbf{(E)} overlap ratio between basins of the top $100$ local optima.} 
    \label{fig:rq2}
\end{figure}


Addressing \textbf{RQ2} is not straightforward, since there exist two intertwined paradigms for identifying basins of a local optimum: the \textit{first-} and \textit{best-}improvement local search. We conducted analyses under both paradigms to see whether top local optima possess larger basins, thereby increasing their likelihood of being encountered during the search process.

\paragraph{Best-improvement basins.} In order to determine the basins of best-improvement for each local optimum, we conducted exhaustive best-improvement local searches starting from every configuration within the landscapes, and recorded the terminating local optimum. For all $32$ landscapes investigated, we observed significant correlation between basin size and the fitness of local optima (see~\pref{fig:rq2_corr}). In particular, local optima with lower fitness have smaller basins, encompassing only a few configurations. On the other hand, those with higher fitness tend to feature significantly larger basins, ranging from hundreds to thousands of configurations (see~\pref{fig:rq2}A). Considering that under best-improvement search, each configuration deterministically converges to a specific basin, the expected basin size for a local optimum in our landscapes is thereby relatively small, often just a few dozen of configurations. This highlights that global peaks' basins are substantially larger than average, increasing their accessibility during optimization processes. Actually, since the basin of each local optimum is exhaustively determined, the likelihood of reaching a given local optimum via a randomly initiated best-improvement search is then directly proportional to its basin size. By plotting the cumulative distribution of local optimum basin sizes, plotted against fitness percentiles, we found that there is a $50\%$ probability of achieving a satisfactory fitness level at the first encounter of a local optimum (see \textcolor{greenstar}{\faIcon[regular]{star}} in~\pref{fig:rq2}A; \textsc{Appendix} $3$). Even more notably, for the top $0.1\%$ of local optima, there remains a competitive probability of reaching them on the first attempt (see \textcolor{purplestar}{\faIcon[regular]{star}} in \pref{fig:rq2}A).

\paragraph{First-improvement basins.} Our previous observations also persist if we consider the first-improvement paradigm. Here, despite significant speed-up strategies in our FLA framework, it is still computationally prohibitive to exhaustively identify the first-improvement basins. Alternatively, we focused on analyzing basins for local optima within the top, middle, and bottom $0.1\%$ fitness percentiles, respectively. The results revealed that the top local optima possess dominantly large basins, with their median encompassing over half of all configurations, significantly surpassing the other two groups (see~\pref{fig:rq2}C, Wilcoxon rank-sum test $p<10^{-i}$). This observation led us to hypothesize substantial overlap among the largest basins. Indeed, we found that the basins of the top $0.1\%$ local optima overlap by an average of $94.4\%\pm2.1\%$ (mean$\pm$standard deviation). Moreover, an inspection of $10^3$ random configurations in each landscape showed that most simultaneously lie within the basins of more than $50\%$ of the local optima (see~\pref{fig:rq2}D). Such overlap can then give rise to optimization contingency, where the same starting configuration might lead to different local optima, not necessarily be the best accessible one. To investigate this, and to verify the accessibility of global peaks, we conducted $10^9$ runs of first-improvement local search across each landscape. We observed that fitter local optima were visited significantly more often (see~\pref{fig:rq2}B; \textsc{Appendix} $3$), with half of the runs reaching local optima within the top $7.4\%$ fitness percentile (\textcolor{purplestar}{\faIcon[regular]{star}}). While notable, these results are less pronounced compared to those from greedy best-improvement search, likely due to the extensive basin overlap~\cite{WhitleyHH13}.

\begin{quotebox}
    \noindent
    \faLightbulb \, \textbf{Response to RQ2:} \textit{Global peaks in software configuration landscapes feature dominant size of basins, which allow them to be accessed with higher chance. }
\end{quotebox}

\subsection{Local Optima Can be Easily Escaped}
\label{sec:rq3}

\begin{figure}[t!]
    \centering
    \includegraphics[width=.6\linewidth]{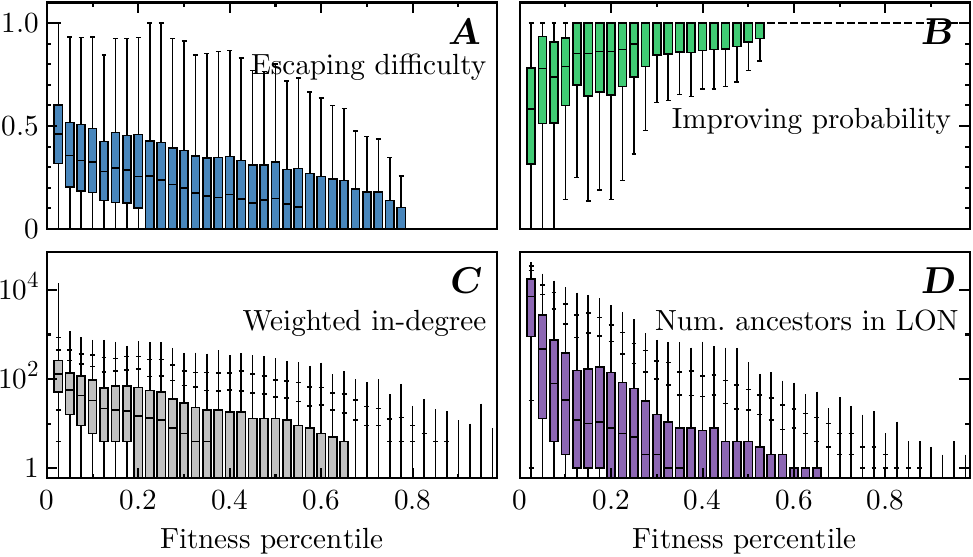}
    \caption{Distribution of local optimum attributes calculated from LONs versus fitness percentile (the lower the better) for \textsc{LLVM}-\texttt{W1}.} 
    \label{fig:rq3}
\end{figure}

\begin{figure}[t!]
    \centering
    \includegraphics[width=.4\linewidth]{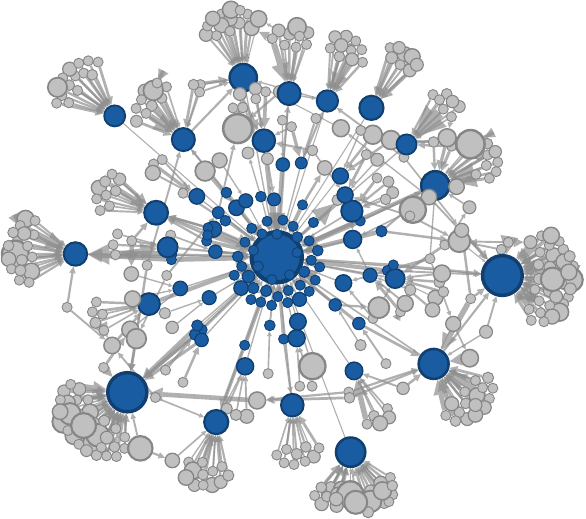}
    \caption{A part of the \textsc{LLVM}-\texttt{W1} LON containing the global peaks. The global optimum and its nearest neighbors are colored in blue, and node radius indicates the basin size.} 
    \label{fig:lon}
\end{figure}

To address \textbf{RQ3}, we need an appropriate model to model the potential connectivity among different local optimum basins. Accordingly, we constructed an LON for each landscape as outlined in~\pref{sec:methods}. Subsequently, we analyzed various features of local optima extracted from these LONs.

\paragraph{Escaping from inferior local optima.} To order to explore how challenging it is to escape from a local optimum in configuration landscapes, we measured the \textit{escaping difficulty} for each local optimum within the LON. This metric is defined by the proportion of $2$-kick perturbations that return to the original local optimum. As shown in~\pref{fig:rq2_corr}, there is a negative correlation between escaping difficulty and the fitness of local optima. Specifically, the least fit local optima can be escaped from using almost any $2$-kick perturbation (see~\pref{fig:rq3}A). We attribute this to their smaller basin sizes, a hypothesis supported by a strong correlation between basin size and escaping difficulty ($0.90\pm0.04$). As local optima become fitter, escaping difficulty increases, yet even near-optimal ones offer a considerable chance of escape, which benefits the exploration phase of optimization algorithms. However, escaping from a local optimum does not always lead to improved fitness. To validate this assertion, we calculated the \textit{improving probability} for each local optimum, defined as the fraction of escape moves leading to a fitter local optimum. Interestingly, this probability also shows a negative correlation with fitness values (\pref{fig:rq2_corr}). While nearly all escape moves from inferior local optima lead to higher fitness, this likelihood decreases for fitter local optima (\pref{fig:rq3}B). Nevertheless, even for top local optima, about half of the escape moves could potentially result in improvements.

\paragraph{Transitions to global peaks.} A pertinent question arising from our analysis is whether such escapability can eventually lead to transitions from local optima to global peaks. To find out, we first examined the \textit{in-degree} of each local optimum within the LON, weighted by the transition frequency. The results indicate that top local optima generally possess higher in-degrees than their inferior counterparts (see~\pref{fig:rq3}C), suggesting a relative ease in transitioning from less optimal regions to more advantageous ones, even with simple $2$-kick perturbations. While this demonstrates direct connections between local optima, it does not capture the entire connectivity landscape. Therefore, we also analyzed the total number of \textit{ancestors} for each local optimum in the LON. This analysis revealed that, on average, $78.4\%\pm16.5\%$ of local optima can transition to the top-$10$ local optima across all scenarios. \pref{fig:lon} depicts the topology of one such LON, illustrating that despite the geographical randomness of local optima, there is a recognized pattern of interconnectedness forming local communities, or landscape \textit{funnels}~\cite{OnuchicW04}. These funnels are typically centered around prominent local optima with large basins, surrounded by numerous less fit local optima capable of transitioning towards them.

To verify these findings, we ran two classic search-based software engineering~\cite{HarmanMZ12} algorithms, particle swarm optimization (PSO) and genetic algorithm (GA), on all the landscapes, with a budget of $2,000$ function evaluations and $100$ repetitions. 
The results reported in~\pref{fig:lon} shows that both algorithms can consistently locate the top $0.02\%$ local optima, with some even approaching the $0.001\%$ level, nearly approaching the global optimum.

\begin{quotebox}
    \noindent
    \faLightbulb \, \textbf{Response to RQ3:} \textit{Despite the ruggedness of software configuration landscapes, inferior local optima can be easily escaped with simple perturbations, and can largely transition to the global peaks.}
\end{quotebox}

\subsection{Landscape Similarities Across Workloads}
\label{sec:rq4}

\begin{figure}[t!]
    \centering
    \includegraphics[width=.6\linewidth]{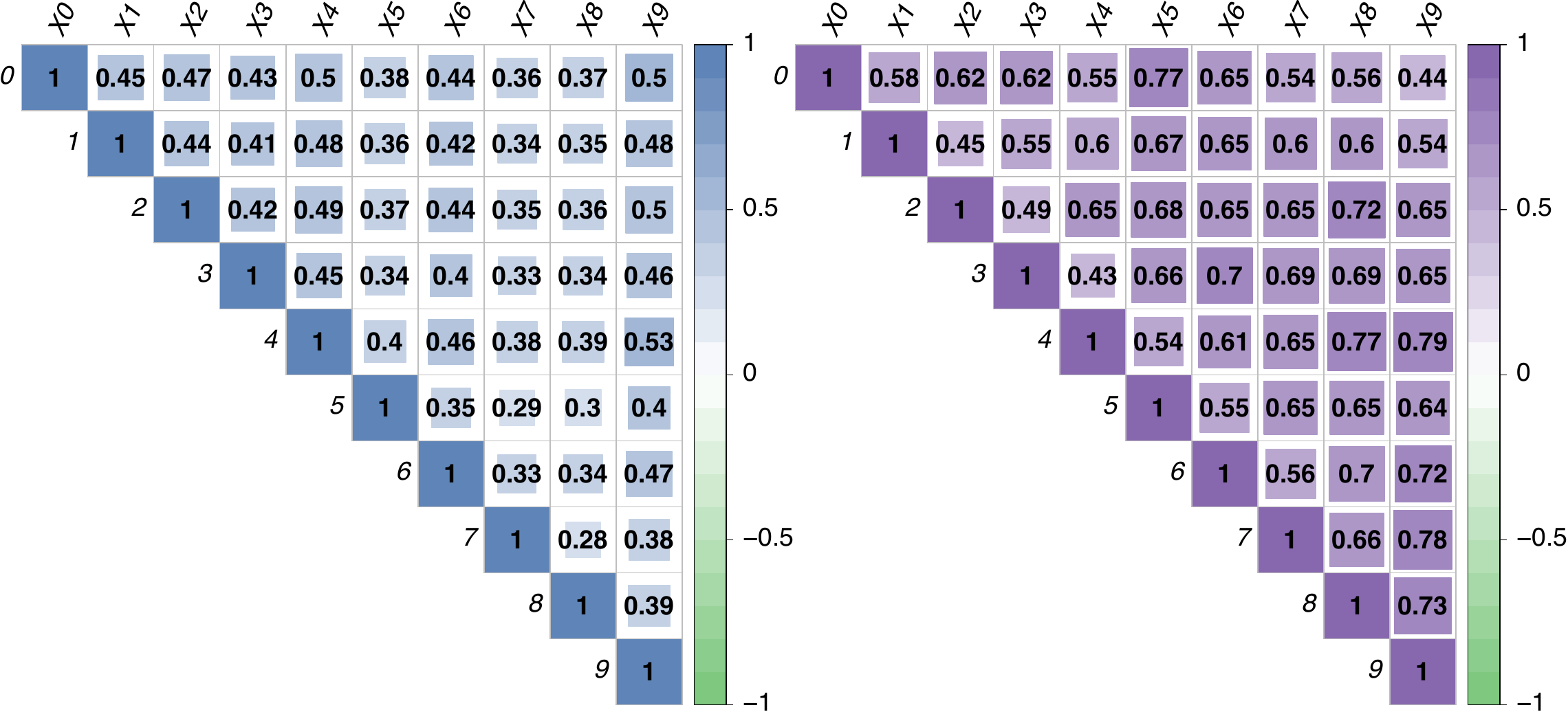}
    \caption{Similarity between the configuration landscapes of \textsc{LLVM} across workloads as measured by (left): Spearman correlation between \texttt{run\_time} of configurations, and (right): structural similarity of the corresponding LONs.}
    \label{fig:rq4_corr}
\end{figure}

\begin{figure}[t!]
    \centering
    \includegraphics[width=.6\linewidth]{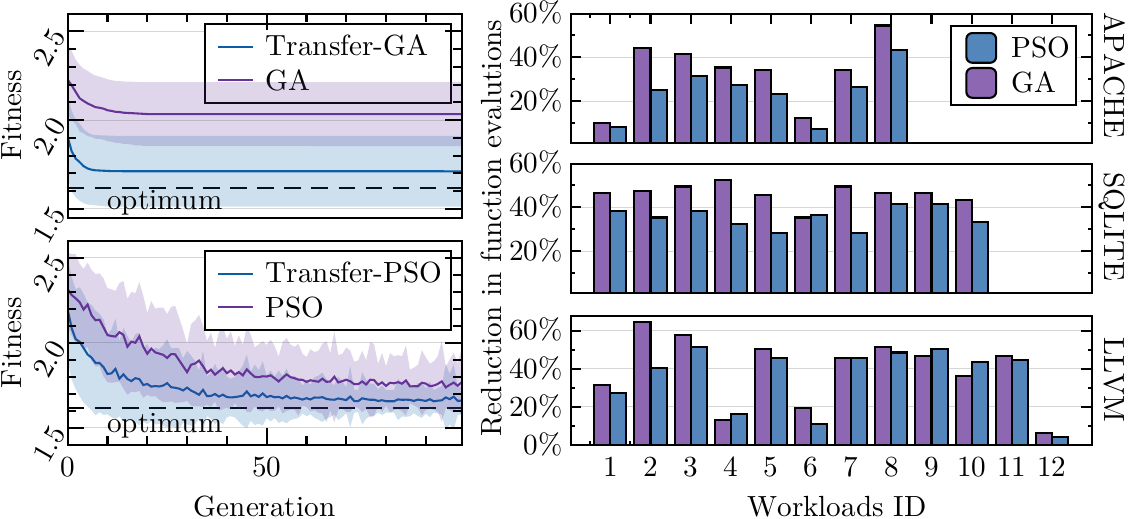}
    \caption{(left): evolution trajectories of warm-started and vanilla version of GA and PSO on \textsc{LLVM}-\texttt{W1}. (right): reduction in function evaluations for warm-started version of GA and PSO to reach the best fitness level achieved by the vanilla version on \textsc{LLVM}-\texttt{W1}.}
    \label{fig:rq4_exp}
\end{figure}

To investigate \textbf{RQ4}, we first assessed the similarity in configuration rankings across different workloads within the same system. We further adopted high-level features learned from LONs as proxies to quantify structural similarities between landscapes of different workloads. Finally, we demonstrated how these identified similarities can be effectively utilized to expedite optimization via warm-start~\cite{FeurerSH15}.

\paragraph{Similarity across workloads.} We first leveraged the Spearman correlation coefficient to quantify the similarity between fitness values of different workloads within each system. This served as a indicator of whether a configuration tends to have similar performance ranks for different workloads. The results show that, while coefficient values can vary, in general a positive correlation exists between fitness values across workloads (see~\pref{fig:rq4_corr} left). In addition, inspired by the recent work~\cite{HuangL23a}, we applied the \texttt{GL2Vec} graph embedding method~\cite{ChenK19} to generate high-level feature representations of each LON, which are able to preserve the intrinsic topology of the network. From the Spearman correlation of these vectors shown in~\pref{fig:rq4_corr} (right), we clearly see that, the LON structure of most workloads are highly similar. This aligns with the fitness correlation, and implies that the connections and clustering of local optima follow similar patterns across various workloads. 

\paragraph{Warm-starting optimizations.} The observed fitness correlation and structural similarity make it natural and appealing to warm-start an optimization process by reusing configurations that are known to perform well on similar tasks. To validate this assertion, we integrated such mechanism into PSO and GA. Specifically, for each workload, we initialized the population using the best configurations found in historical optimizations of the $3$ most similar tasks. In particular, the historical optimization data were readily available from \textbf{RQ3}, and we leverage the structural similarities between LONs as the measure of task similarity. We ran all both warm-started and vanilla version of PSO and GA on each workload for $100$ times to ensure statistically robust conclusions. As depicted in the left panel of~\pref{fig:rq4_exp} (full results are in \textsc{Appendix} $4$), we found that the warm-started version of both algorithms can outperform the vanilla version in most cases. More importantly, by employing warm-start, we typically need considerably fewer function evaluations to reach the best fitness level achieved by the vanilla version---As can be seen from~\pref{fig:rq4_exp} (right), in most scenarios the warm-started algorithms can save over $20\%$ function evalutions, and in some cases even up to $60\%$. 


\begin{quotebox}
    \noindent
    \faLightbulb \, \textbf{Response to RQ4:} \textit{There exists structural similarity between configuration landscapes of different workloads, which can be exploited to expedit optimization via reusing prominent configurations in similar tasks for warm-start.}
\end{quotebox}


\section{Related Work}
\label{sec:related_works}

\paragraph{Software configuration tuning.} Efficiently tunning of configurable software systems has been a long-desired goal in software engineering. Although a plethora of techniques have been developed to automate this laborious process, e.g., random search~\cite{OhBMS17}, hill-climbing~\cite{LiZMTZBF14}, genetic algorithms~\cite{ShahbazianKBM20}, and sequential model-based ones~\cite{Nair0MSA20}, they are usually based on intuition or certain assumption of the configuration space. Consequently, they may not be sufficiently efficient due to the lack of knowledge for the underlying system.

\paragraph{Performance analysis on software systems.} Traditionally, the configuration-performance interaction has been explored by \textit{black-box} approaches like performance-influence modeling~\cite{OlaecheaRGC14,KalteneckerGSGA19,SiegmundGAK15}, where the goal is to model such interaction using representative samples drawn from the system under investigation. Another line of related research use machine learning methods to predict the performance of a given configuration~\cite{HaZ19,ValovGC15,Nair0MSA20}. While our collected data can also be used for the above purposes, we note that they are however, orthogonal and complementary to this study. For example, \cite{JamshidiSVKPA17} found that performance models can be linearly transferred between workloads, which aligns with our results for \textbf{RQ4}, but we explore this from a landscape topology perspective. On the flip side, there is a growing interest in applying \textit{white-box} approaches to explore additional insights of the interactions between configuration options versus performance, e.g.,~\cite{VelezJSSAK20,VelezJSAK21}. Particularly, the white-box methods focus on interogatting the system's source codes to identify the regions of a system responsible for the performance difference among configurations, which is also orthogonal to this present work.

\paragraph{Landscape analysis for software system.} Landscape analysis has been a subject of rigorous research within the meta-heuristics community for several decades, but is a fairly new topic in software engineering. Some workds have brifiely discussed the local optima in software configuration~\cite{JamshidiVKS18,DoncktWQDM20}, but they achieved this only with naive 2/3D projections. So far there is no dedicated work on comprehensively investigating the topography of configuration landscapes, and this work is the first to do so.

\vspace{-.3em}
\section{Conclusion and Discussions}
\label{sec:conclusion}


This work conducted the first FLA on $32$ configuration landscapes of three real-world software systems. We observed a fairly rugged topography for these landscapes with abundant pervasive local optima (\textbf{RQ1}), which can potentially present challenge for not only navigation, but also performance modeling, as have been known in the protein engineering community~\cite{FahlbergFHR23}. Intriguingly, we later observed that albeit their rugged surface, configuration landscapes are highly navigable in the sense that global peaks often feature dominantly large basins of attraction (\textbf{RQ2}), and those inferior ones can be easily escaped with simple perturbations (\textbf{RQ3}). These then highlight the need for efficient exploration and escaping mechanisms for searching. Finally, we discovered structural similarity among configuration landscapes across workloads (\textbf{RQ4}), providing concrete evidence to advocate configuration reuse strategies in practice, which has attracted attention in the community~\cite{KinneerGG21}.

There are far more interesting properties regarding configuration landscape that have yet to be explored in the future. We hope that this work will be the first of a continually growing sequence of rigorous research for configuration landscape analysis. Additionally, in open-sourcing the developed FLA framework and gathered performance data, we also hope to make landscape analysis and software configuration research more accessible and reproducible by the community.

\section*{Acknowledgment}
This work was supported in part by the UKRI Future Leaders Fellowship under Grant MR/S017062/1 and MR/X011135/1; in part by NSFC under Grant 62376056 and 62076056; in part by the Royal Society under Grant IES/R2/212077; in part by the EPSRC under Grant 2404317; in part by the Kan Tong Po Fellowship (KTP\textbackslash R1\textbackslash 231017); and in part by the Amazon Research Award and Alan Turing Fellowship.

\bibliographystyle{IEEEtran}
\bibliography{lon}

\newpage
\appendix
\section{Data Collection}
\label{sec:system}

\subsection{Configurable Software Systems}
\label{sec:csc}

To ensure the practicality and generality of our empirical findings, this paper considers investigating three widely used configurable software systems with diverse engineering functionalities, including compiler, Web server, and database management system. In the following paragraphs, we outline their key characteristics including the engineering narration, configuration options considered in our experiments, and the settings of workloads.

\begin{itemize}
    \item\textsc{\textbf{LLVM}}\footnote{https://llvm.org/}: The \textsc{LLVM} Project is a collection of modular compiler and toolchain technologies. It provides a modern, SSA-based compilation strategy that supports both static and dynamic compilation of any programming language. $\blacktriangleright$ \textsc{LLVM} has more than $578$ configuration options while we choose $20$ of them for our empirical study. $\blacktriangleright$ $12$ test suites from the widely used  \texttt{PolyBench} benchmark suite\footnote{http://web.cse.ohio-state.edu/~pouchet.2/software/polybench/} are chosen to constitute the workloads. $\blacktriangleright$ The \texttt{run\_time} for the compiled program is used as the fitness function to measure the quality of a configuration of the \texttt{LLVM}.

    \item\textsc{\textbf{Apache}}\footnote{https://httpd.apache.org/}: The \textsc{Apache} HTTP Server Project aims to provide a robust and scalable HTTP service. $\blacktriangleright$ It consists of multiple modules, the core of which has $89$ configuration options and $21$ configuration options for MPM module. Here we choose $15$ options directly related to the quality of a configuration in our experiments. $\blacktriangleright$ $9$ different running environments are generated by using the Apache HTTP server benchmarking tool\footnote{https://httpd.apache.org/docs/2.4/programs/ab.html}. $\blacktriangleright$ We use the request handled per second as the fitness function.

    \item\textsc{\textbf{SQLite}}\footnote{https://www.sqlite.org/index.html}: This is an embedded database project. Instead of maintaining a separate server process, \textsc{SQLite} directly reads and writes data to disk. $\blacktriangleright$ It has $50$ compile-time and $29$ run-time configuration options and we chose $18$ of them in this study. $\blacktriangleright$ We used the \textsc{SQLite} Benchmark\footnote{https://github.com/ukontainer/sqlite-bench} to constitute $10$ different running workloads. $\blacktriangleright$ The writing speed in sequential key order in async mode (\texttt{fillseqsync}) is used as the fitness function.
\end{itemize}

The meta information of the selected configuration options (parameters) for these systems are listed in Tables~\ref{tab:llvm_parameters} to \ref{tab:sqlite_parameters}. The settings of different workloads for each system are listed in~\pref{tab:environment_lookup}. In the following paragraphs, we introduce the corresponding attributes for different workloads.
\begin{itemize}
    \item For the \textsc{LLVM}, we adopted different compiling file to constitute different workloads, as indicated by the attribute \texttt{program\_name}. 
    \item For the \textsc{Apache}, there are two attributes to setup the system, and their different combinations constitute different workloads:
        \begin{itemize}
            \item \texttt{requests} represents the number of requests to perform for the benchmarking session. The default is to just perform a single request which usually leads to non-representative benchmarking results.
            \item \texttt{concurrency} represents the number of multiple requests to perform concurrently. The default is one request at a time.
        \end{itemize}
    \item For the \texttt{SQLITE}, the worklodas are based on two system attributes:
        \begin{itemize}
            \item{\texttt{num}} indicates the number of entries.
            \item{\texttt{value\_size}} represents the value size.
        \end{itemize}
\end{itemize}

\begin{table}[h!]
    \centering
    \small
    \caption{Selected configuration options for \textsc{LLVM}}
    \label{tab:llvm_parameters}
    \begin{tabular}{||ccc||}
        \hline
        Index & Parameter & Value \\ \hline \hline
        $1$ & \texttt{inline} &$\{\texttt{on},\texttt{off}\}$ \\
        $2$ & \texttt{openmpopt} & $\{\texttt{on},\texttt{off}\}$ \\
        $3$ & \texttt{mldst-motion} & $\{\texttt{on},\texttt{off}\}$ \\
        $4$ & \texttt{gvn} &$\{\texttt{on},\texttt{off}\}$ \\
        $5$ & \texttt{jump-threading} &$\{\texttt{on},\texttt{off}\}$ \\
        $6$ & \texttt{correlated-propagation} &$\{\texttt{on},\texttt{off}\}$ \\
        $7$ & \texttt{elim-avail-extern} &$\{\texttt{on},\texttt{off}\}$ \\
        $8$ & \texttt{tailcallelim} &$\{\texttt{on},\texttt{off}\}$ \\
        $9$ & \texttt{constmerge} &$\{\texttt{on},\texttt{off}\}$ \\
        $10$ & \texttt{dse} &$\{\texttt{on},\texttt{off}\}$ \\
        $11$ & \texttt{slp-vectorizer} &$\{\texttt{on},\texttt{off}\}$ \\
        $12$ & \texttt{callsite-splitting} &$\{\texttt{on},\texttt{off}\}$ \\
        $13$ & \texttt{argpromotion} &$\{\texttt{on},\texttt{off}\}$ \\
        $14$ & \texttt{aggressive-instcombine} &$\{\texttt{on},\texttt{off}\}$ \\
        $15$ & \texttt{polly-simplify} &$\{\texttt{on},\texttt{off}\}$ \\
        $16$ & \texttt{polly-dce} &$\{\texttt{on},\texttt{off}\}$ \\
        $17$ & \texttt{polly-optree} &$\{\texttt{on},\texttt{off}\}$ \\
        $18$ & \texttt{polly-delicm} &$\{\texttt{on},\texttt{off}\}$ \\
        $19$ & \texttt{polly-opt-isl} &$\{\texttt{on},\texttt{off}\}$ \\
        $20$ & \texttt{polly-prune-unprofitable} &$\{\texttt{on},\texttt{off}\}$ \\ \hline
    \end{tabular}
  \end{table}

  \begin{table}[h!]
    \centering
    \small
    \caption{Selected configuration options for \textsc{Apache}}
    \label{tab:apache_parameters}
    \begin{tabular}{||ccc||}
        \hline
        Index & Parameter & Value \\ \hline \hline
        $1$ & \texttt{AcceptFilter} & $\{\texttt{nntp},\texttt{http}\}$ \\
        $2$ & \texttt{KeepAlive} & $\{\texttt{on},\texttt{off}\}$ \\
        $3$ & \texttt{KeepAliveTimeout} & $\{1,\ldots,300\}$ \\
        $4$ & \texttt{MaxKeepAliveRequests} & $\{1,\ldots,2^{10}\}$ \\
        $5$ & \texttt{TimeOut} & $\{1,\ldots,300\}$ \\
        $6$ & \texttt{MaxConnectionsPerChild} & $\{1,\ldots,1,000\}$ \\
        $7$ & \texttt{MaxMemFree} & $\{2^{10},\ldots,2^{20}\}$ \\
        $8$ & \texttt{MaxRequestWorkers} & $\{100,\ldots,3,000\}$ \\
        $9$ & \texttt{MaxSpareThreads} & $\{50,\ldots,500\}$ \\
        $10$ & \texttt{MinSpareThreads} & $\{20,\ldots,250\}$ \\
        $11$ & \texttt{SendBufferSize} & $\{2^{10},\ldots,2^{16}\}$ \\
        $12$ & \texttt{ServerLimit} & $\{100,\ldots,3,000\}$ \\
        $13$ & \texttt{StartServers} & $\{1,\ldots,10\}$ \\
        $14$ & \texttt{ThreadLimit} & $\{10,\ldots,200\}$ \\
        $15$ & \texttt{ThreadsPerChild} & $\{10,\ldots,200\}$ \\ \hline
    \end{tabular}
    \end{table}

  \begin{table}[h!]
    \centering
    \tiny
    \small
    \caption{Selected configuration options for \textsc{SQLite}}
    \label{tab:sqlite_parameters}
    \begin{tabular}{||ccc||}
        \hline
        Index & Parameter & Value \\ \hline \hline
        $1$ & \texttt{SQLITE\_SECURE\_DELETE} & $\{\texttt{on},\texttt{off}\}$ \\
        $2$ & \texttt{SQLITE\_TEMP\_STORE} & $\{0,1,2,3\}$ \\
        $3$ & \texttt{SQLITE\_ENABLE\_AUTO\_WRITE} &$\{\texttt{on},\texttt{off}\}$ \\
        $4$ & \texttt{SQLITE\_ENABLE\_STAT3} &$\{\texttt{on},\texttt{off}\}$ \\
        $5$ & \texttt{SQLITE\_DISABLE\_LFS} &$\{\texttt{on},\texttt{off}\}$ \\
        $6$ & \texttt{SQLITE\_OMIT\_AUTO\_INDEX} &$\{\texttt{on},\texttt{off}\}$ \\
        $7$ & \texttt{SQLITE\_OMIT\_BETWEEN\_OPT} &$\{\texttt{on},\texttt{off}\}$ \\
        $8$ & \texttt{SQLITE\_OMIT\_BTREECOUNT} &$\{\texttt{on},\texttt{off}\}$ \\
        $9$ & \texttt{SQLITE\_OMIT\_LIKE\_OPT} &$\{\texttt{on},\texttt{off}\}$ \\
        $10$ & \texttt{SQLITE\_OMIT\_LOOKASIDE} &$\{\texttt{on},\texttt{off}\}$ \\
        $11$ & \texttt{SQLITE\_OMIT\_OR\_OPT} &$\{\texttt{on},\texttt{off}\}$ \\
        $12$ & \texttt{SQLITE\_OMIT\_QUICKBALANCE} &$\{\texttt{on},\texttt{off}\}$ \\
        $13$ & \texttt{SQLITE\_OMIT\_SHARED\_CACHE} &$\{\texttt{on},\texttt{off}\}$ \\
        $14$ & \texttt{CacheSize} & $\{1,\ldots,10,240\}$ \\
        $15$ & \texttt{AutoVacuumON} & $\{0,1,2\}$ \\
        $16$ & \texttt{ExclusiveLock} &$\{\texttt{on},\texttt{off}\}$ \\
        $17$ & \texttt{PageSize} & $\{1,\ldots,10,240\}$ \\
        $18$ & \texttt{Wal} &$\{\texttt{on},\texttt{off}\}$ \\ \hline
    \end{tabular}
  \end{table}

  \begin{table*}[t!]
    \centering
    \small
    \caption{Lookup table of settings of different running environments for three configurable software systems.}
    \label{tab:environment_lookup}
    \begin{tabular}{||c|c|cccc}
        \hline
        Sys.& \textsc{LLVM}             & \multicolumn{2}{c|}{\textsc{SQLite}}                                           & \multicolumn{2}{c||}{\textsc{Apache}}                                                        \\ \hline \hline
        \multicolumn{1}{||c|}{Index} & \texttt{program\_name} & \multicolumn{1}{c}{\texttt{num}} & \multicolumn{1}{c|}{\texttt{value\_size}} & \multicolumn{1}{c}{\texttt{requests}} & \multicolumn{1}{c||}{\texttt{concurrency}}  \\ \hline
        
        \multicolumn{1}{||c|}{$1$}     & 2mm              & \multicolumn{1}{c}{$10$}             & \multicolumn{1}{c|}{$100$}        & \multicolumn{1}{c}{$50$}                & \multicolumn{1}{c||}{$50$}                      \\ 
        \multicolumn{1}{||c|}{$2$}     & 3mm              & \multicolumn{1}{c}{$10$}             & \multicolumn{1}{c|}{$1,000$}       & \multicolumn{1}{c}{$100$}               & \multicolumn{1}{c||}{$100$}                     \\ 
        \multicolumn{1}{||c|}{$3$}     & atax             & \multicolumn{1}{c}{$10$}             & \multicolumn{1}{c|}{$10,000$}      & \multicolumn{1}{c}{$100$}               & \multicolumn{1}{c||}{$100$}                 \\ 
        \multicolumn{1}{||c|}{$4$}     & correlation      & \multicolumn{1}{c}{$10$}             & \multicolumn{1}{c|}{$30,000$}      & \multicolumn{1}{c}{$200$}               & \multicolumn{1}{c||}{$200$}                   \\ 
        \multicolumn{1}{||c|}{$5$}     & covariance       & \multicolumn{1}{c}{$100$}            & \multicolumn{1}{c|}{$100$}        & \multicolumn{1}{c}{$250$}               & \multicolumn{1}{c||}{$250$}                 \\ 
        \multicolumn{1}{||c|}{$6$}     & deriche          & \multicolumn{1}{c}{$100$}            & \multicolumn{1}{c|}{$100$}        & \multicolumn{1}{c}{$300$}               & \multicolumn{1}{c||}{$300$}               \\ 
        \multicolumn{1}{||c|}{$7$}     & doitgen          & \multicolumn{1}{c}{$100$}            & \multicolumn{1}{c|}{$1,000$}       & \multicolumn{1}{c}{$400$}               & \multicolumn{1}{c||}{$400$}                    \\ 
        \multicolumn{1}{||c|}{$8$}     & fdtd2d           & \multicolumn{1}{c}{$100$}            & \multicolumn{1}{c|}{$10,000$}      & \multicolumn{1}{c}{$500$}               & \multicolumn{1}{c||}{$500$}                    \\ 
        \multicolumn{1}{||c|}{$9$}     & gemm             & \multicolumn{1}{c}{$100$}            & \multicolumn{1}{c|}{$30,000$}      & \multicolumn{1}{c}{$1,000$}              & \multicolumn{1}{c||}{$100$}                \\ \cline{5-6}
        \multicolumn{1}{||c|}{$10$}    & symm             & \multicolumn{1}{c}{$1,000$}           & \multicolumn{1}{c|}{$10$}         &                                                              \\ \cline{3-4} 
        \multicolumn{1}{||c|}{$11$}    & syr2k            &                                     &                                 &                                        &                                                                   \\ 
        \multicolumn{1}{||c|}{$12$}    & syrk             &                                     &                                 &                                        &                                                   \\ 
        \multicolumn{1}{||c|}{$13$}    & trmm             &                                     &                                 &                                        &                                                   \\ \cline{1-2}  
    \end{tabular}
\end{table*}

\subsection{Summary of our computational resrouces}
\label{sec:resources}

All of our data collection experiments were run on a cluster with $20$ nodes, each of which is equiped with Intel$^{\text{\textregistered}}$ Core$^{\text{TM}}$ i$7$--$8700$ CPU@$3.10$GHz and 16GB memory. Evaluating all $86$M configurations from the $3$ systems with $5$ repetitions took about $6$ months to complete, which results in a total of more than $86,400$ CPU hours. For the landscape construction and analyses, all the experiments were carried out using a single node with Intel$^{\text{\textregistered}}$ Xeon$^{\text{\textregistered}}$ Platinum $8260$ CPU@$2.40$GHz and $256$GB memory.

\section{Fitness Landscape Analysis}
\label{sec:graph}

By representing the software configuration landscape as a directed graph\footnote{Implemented using \texttt{NetworkX} package: https://networkx.org/.}, many classic fitness landscape analysis (FLA) methods can be implemented in straightforward graph traversal manners. Here, we delineate the essential ideas and implementations of several FLA methods used in this paper. 

\paragraph{Local optima.} A local optimum is a configuration that has no superior neighbor. Once the  landscape is represented as a graph, the local optima can be easily identified by finding the nodes with no outgoing edges, i.e., the \textit{sink} nodes. 

\paragraph{Basin of attraction.} While a rugged landscape can be difficult to optimize due to the pressence of various local optima, not all are equal in terms of the capability of trapping a solver. For a 2D minimization case, this can be envisioned by the fact that each local optimum is located at the bottom of a `basin' in the landscape surface. Configurations in each basin would eventually fall into the corresponding basin bottom, i.e., the local optimum, when following a simplest hill-climbing local search. To determine the basin of attraction of each local optimum in the landscape, we consider two most popular local search paradigms:

\begin{itemize}
    \item \textbf{Best-improvement local search} (Algorithm~\ref{alg:best_impr_search}): In each iteration, the search moves to the neighbor with the highest fitness value. It terminates when no neighbor has a higher fitness value than the current configuration (i.e., a local optimum). For a graph-based landscape, this can be achieved by iteratively selecting the best \textit{successor} of each node until a local optimum is encountered. The \textit{best improvement basin} of a local optimum can be then determined by exhaustively perform such search from each configuration in the landscape, and collect all the configurations that fall into the same local optimum. Note that while this sounds like a computationally expensive task, in practice for landscapes with even millions of configurations, it would take only a few seconds to determine the basin of attraction of each local optimum.
    \item \textbf{First-improvement local search} (Algorithm~\ref{alg:first_impr_search}): Here in each iteration, instead of selecting the best neighbor, the search moves to the first neighbor that it encounters with a higher fitness value. This is implemented by iteratively random selecting a successor of each node until a local optimum is reached. Under this paradigm, identifying the basin of attraction of each local optimum is equal to finding the \textit{ancestors} of a node. 
\end{itemize}

\paragraph{Autocorrelation.} This is a widely used metric for characterizing the ruggedness of a landscape. As briefly introduced in the main paper, it is the autocorrelation $\rho_a$ of a consecutive series of fitness values $\{f_1,\dots,f_n\}$ obtained from a random walk on the landscape. Due to the graph representation of the landscape, performing a random walk on the landscape is equivalent to that on a graph, which can be executed in a lightning fast manner in \texttt{NetworkX}.

\paragraph{Graph embedding.} In this paper, we adopted \texttt{GL2Vec}, an improved version of \texttt{Graph2Vec} to extract low-dimensional features from the LON of each landscape. The generated features are able to capture the topological structure of the LON, and thereby the distribution and connectivity pattern of local optima in the landscape. We employed the implementation of \texttt{GL2Vec} from the \href{https://github.com/benedekrozemberczki/karateclub}{\texttt{Karateclub}} package. 

\begin{algorithm}[t!]
    \caption{Best-Improvement Local Search}
    \label{alg:best_impr_search}
    \KwIn{A starting configuration $\mathbf{c}$; A neighborhood function $\mathcal{N}$; A fitness function $f$}
    \KwOut{A local optima configuration $\mathbf{c}^\ell$}
    \While{$\mathbf{c}$ is not a local optimum}{
        
        $\mathbf{c}^{\prime\star}$ = $\argmax_{\mathbf{c}^\prime \in \mathcal{N}(\mathbf{c})}(f(\mathbf{c}^\prime))$ \;
        \If {$f(\mathbf{c}^{\prime\star}) > f(\mathbf{c})$}{
            $\mathbf{c} \gets \mathbf{c}^{\prime\star}$\;
        }
        \Else{
            $\mathbf{c}$ is a local optimum\;
            \textbf{break}\;
        }
        
    }
\end{algorithm}

\begin{algorithm}[t!]
    \caption{First-Improvement Local Search}
    \label{alg:first_impr_search}
    \KwIn{A starting configuration $\mathbf{c}$; A neighborhood function $\mathcal{N}$; A fitness function $f$}
    \KwOut{A local optima configuration $\mathbf{c}^\ell$}
    \While{$\mathbf{c}$ is not a local optimum}{
        \texttt{Improve} = \texttt{False}\;
        \For{$\mathbf{c}^\prime \in \mathcal{N}(\mathbf{c})$}{
            \If {$f(\mathbf{c}^\prime) > f(\mathbf{c})$}{
            $\mathbf{c} \gets \mathbf{c}^\prime$\;
            \texttt{Improve} = \texttt{True}\;
            \textbf{break}\;
        }
        }
        \If{not \texttt{Improve}}{
        $\mathbf{c}$ is a local optimum\;
        \textbf{break}\;
    }
    }
\end{algorithm}

\section{Full Results for Sections 4.1 and 4.2}
\label{sec:landscape}

\begin{figure*}
    \centering
    \includegraphics[width=1\linewidth]{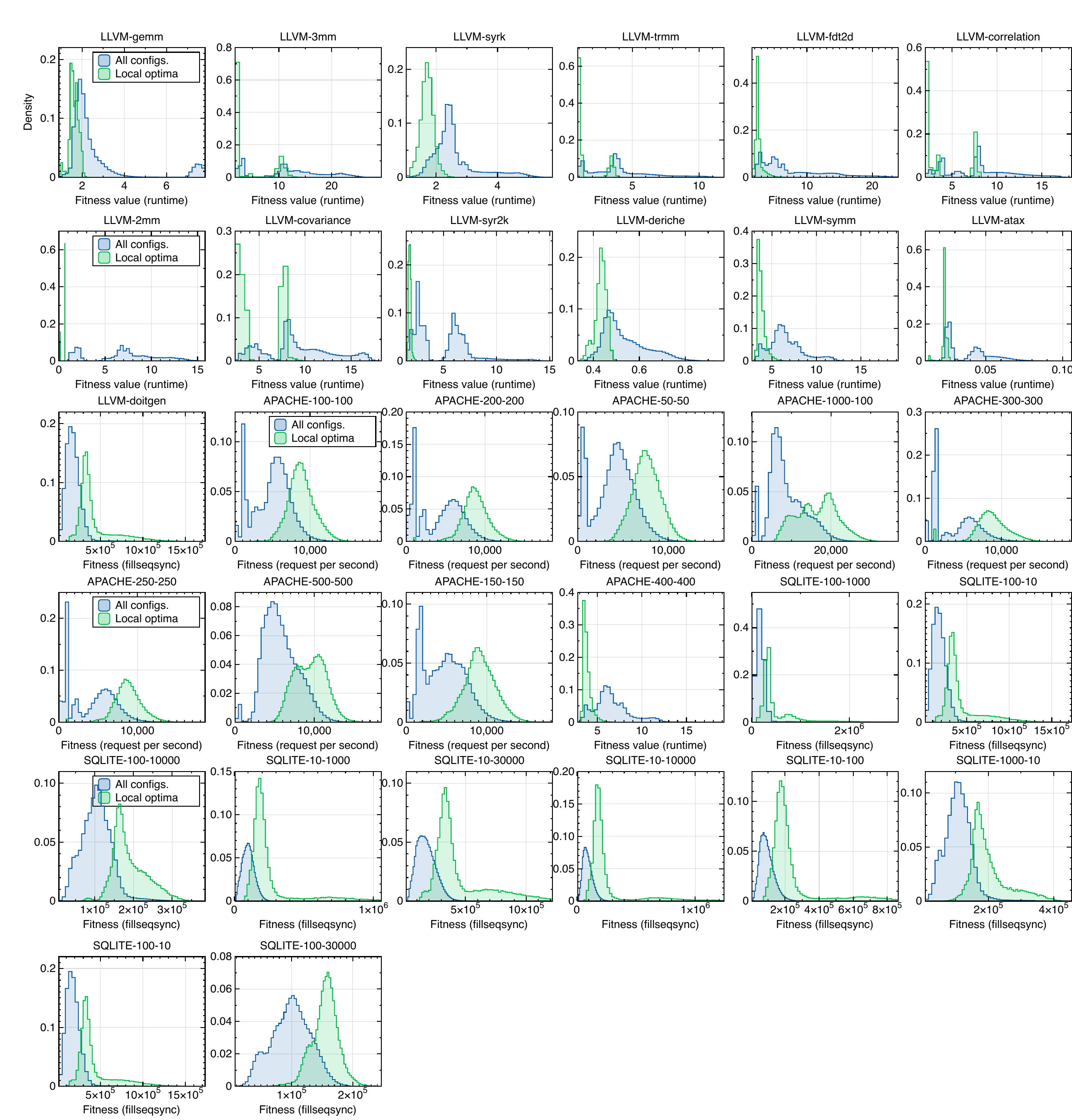}
    \caption{Comaprison of fitness distribution of all configurations (\textcolor{f1blue}{blue}) versus local optima (\textcolor{f1green}{green}) for all studied landscapes. Note that while the objective function for \textsc{LLVM} is minimized, \textsc{Apache} and \textsc{SQLite} have maximized objectives.}
    \label{fig:fit_distribution}
\end{figure*}

\begin{figure*}
    \centering
    \includegraphics[width=1\linewidth]{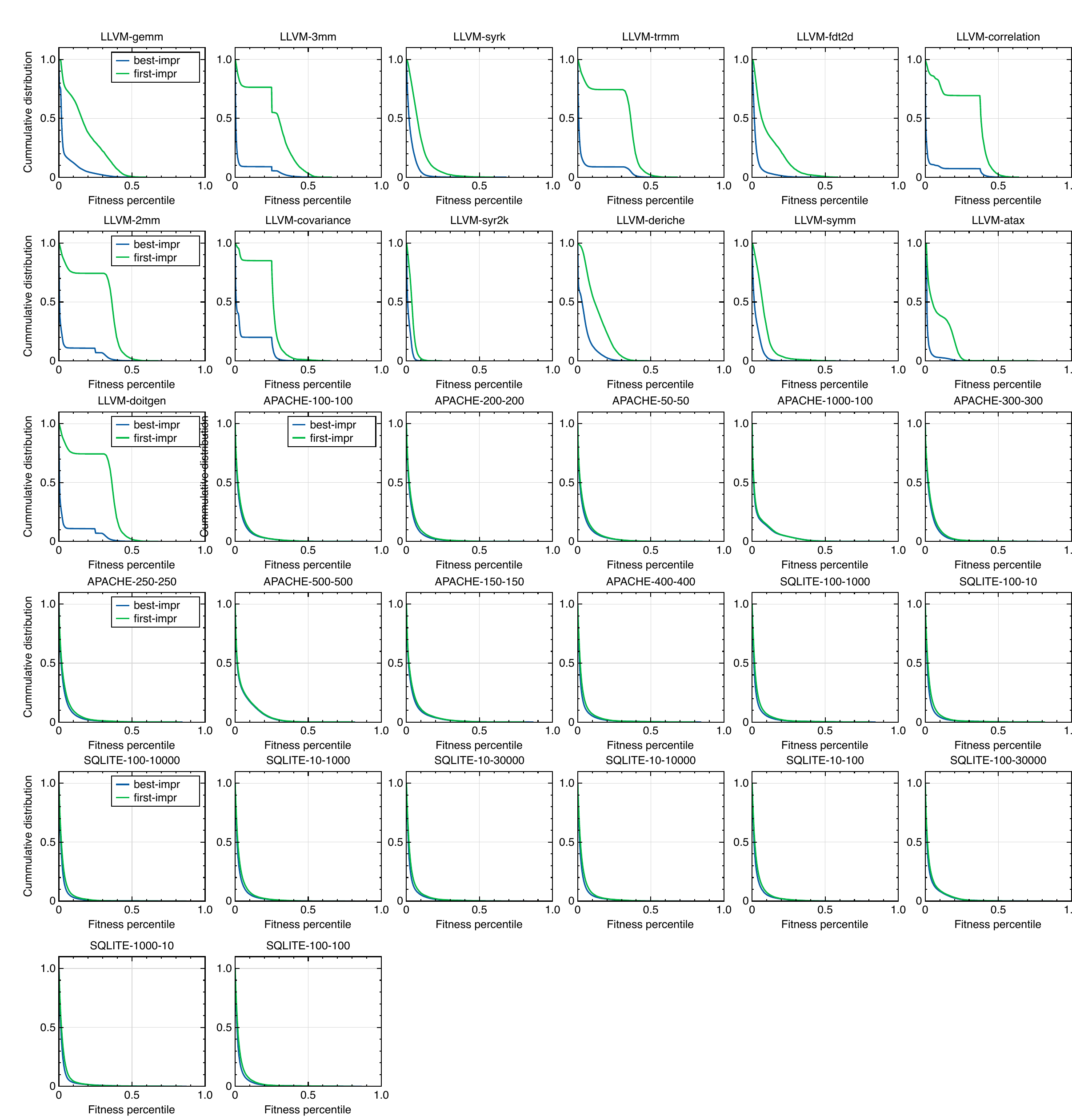}
    \caption{This plot provides the result for the Figure 3(A) and (B) (in the main text) across all scenarios, showing the cummlative distribution of basin size versus the fitness percentile of local optima under both best- and first-improvement local search. Note that for best-improvement, the basin size can be deterministically calculated, and $y=1.0$ indicates the total sum of all basin sizes, which equals to the total number of configurations in the landscape. For first-improvment, the curves are approxiated by conducting $10^9$ runs of randomized local search on the landscape, and hence $y=1.0$ represents the total frequency of visits (i.e., $10^9$).}
    \label{fig:basin}
\end{figure*}

The full results related to Sections $4.1$ to $4.2$ can be found in \pref{tab:meta}, \pref{fig:fit_distribution}, and \pref{fig:basin}.

\begin{table}[t!]
    \caption{Full results related to Figure 1 in tabular form, including the number of local optima in each landscape, and the statistics and $p$-value for comparing the distribution of local optima with random configurations sampled from each landscape.}
    \centering
    \small
    \begin{tabular}{||ll|r|cl||}
        \hline
        System & Workload & $n$ peaks & Stat. & $p$-value \\
        \hline
        \hline
        \textsc{LLVM}     & gemm         & $25,974$ & $0.100$ & $7.6\text{e}^{-6}$ \\
                          & 3mm          & $17,918$ & $0.145$ & $4.6\text{e}^{-2}$ \\
                          & syrk         & $43,522$ & $0.095$ & $2.9\text{e}^{-4}$ \\
                          & trmm         & $24,658$ & $0.092$ & $1.7\text{e}^{-4}$ \\
                          & fdtd2d       & $39,012$ & $0.092$ & $1.7\text{e}^{-4}$ \\
                          & correlation  & $12,782$ & $0.145$ & $4.6\text{e}^{-2}$ \\
                          & 2mm          & $17,494$ & $0.145$ & $4.6\text{e}^{-2}$ \\
                          & covariance   & $20,495$ & $0.145$ & $4.6\text{e}^{-2}$ \\
                          & syr2k        & $24,619$ & $0.139$ & $2.5\text{e}^{-2}$ \\
                          & deriche      & $36,631$ & $0.092$ & $1.7\text{e}^{-4}$ \\
                          & doitgen      & $32,148$ & $0.092$ & $1.7\text{e}^{-4}$ \\
                          & symm         & $42,962$ & $0.095$ & $2.9\text{e}^{-4}$ \\
                          & atax         & $15,388$ & $0.092$ & $1.7\text{e}^{-4}$ \\
        \hline
        \hline
        \textsc{Apache}   & 100\_100     & $204,372$ & $0.067$ & $9.4\text{e}^{-6}$ \\
                          & 200\_200     & $184,939$ & $0.067$ & $9.4\text{e}^{-6}$ \\
                          & 50\_50       & $186,086$ & $0.067$ & $9.4\text{e}^{-6}$ \\
                          & 1000\_100    & $121,153$ & $0.067$ & $9.4\text{e}^{-6}$ \\
                          & 300\_300     & $165,537$ & $0.067$ & $9.4\text{e}^{-6}$ \\
                          & 150\_150     & $179,419$ & $0.067$ & $9.4\text{e}^{-6}$ \\ 
                          & 250\_250     & $177,716$ & $0.067$ & $9.4\text{e}^{-6}$ \\
                          & 400\_400     & $191,534$ & $0.067$ & $9.4\text{e}^{-6}$ \\ 
                          & 500\_500     & $128,825$ & $0.067$ & $9.4\text{e}^{-6}$ \\ 
        \hline
        \hline
        \textsc{SQLite}   & 1000\_10     & $74,261 $ & $0.056$ & $3.2\text{e}^{-18}$\\ 
                          & 100\_100     & $75,324 $ & $0.056$ & $3.2\text{e}^{-18}$\\
                          & 100\_1000    & $72,908 $ & $0.056$ & $3.2\text{e}^{-18}$\\
                          & 100\_10      & $77,017 $ & $0.056$ & $3.2\text{e}^{-18}$\\
                          & 100\_10000   & $71,053 $ & $0.111$ & $2.9\text{e}^{-5}$ \\
                          & 10\_1000     & $79,035 $ & $0.111$ & $2.9\text{e}^{-5}$ \\
                          & 10\_30000    & $75,645 $ & $0.056$ & $3.2\text{e}^{-18}$\\
                          & 10\_10000    & $79,438 $ & $0.056$ & $3.2\text{e}^{-18}$\\
                          & 10\_100      & $78,869 $ & $0.056$ & $3.2\text{e}^{-18}$\\
                          & 100\_30000   & $66,680 $ & $0.111$ & $2.9\text{e}^{-5}$\\                         
        \hline
    \end{tabular}
    \label{tab:meta}
\end{table}

\section{Full Results for Sections 4.3 and 4.4}

\begin{figure*}
    \centering
    \includegraphics[width=1.\linewidth]{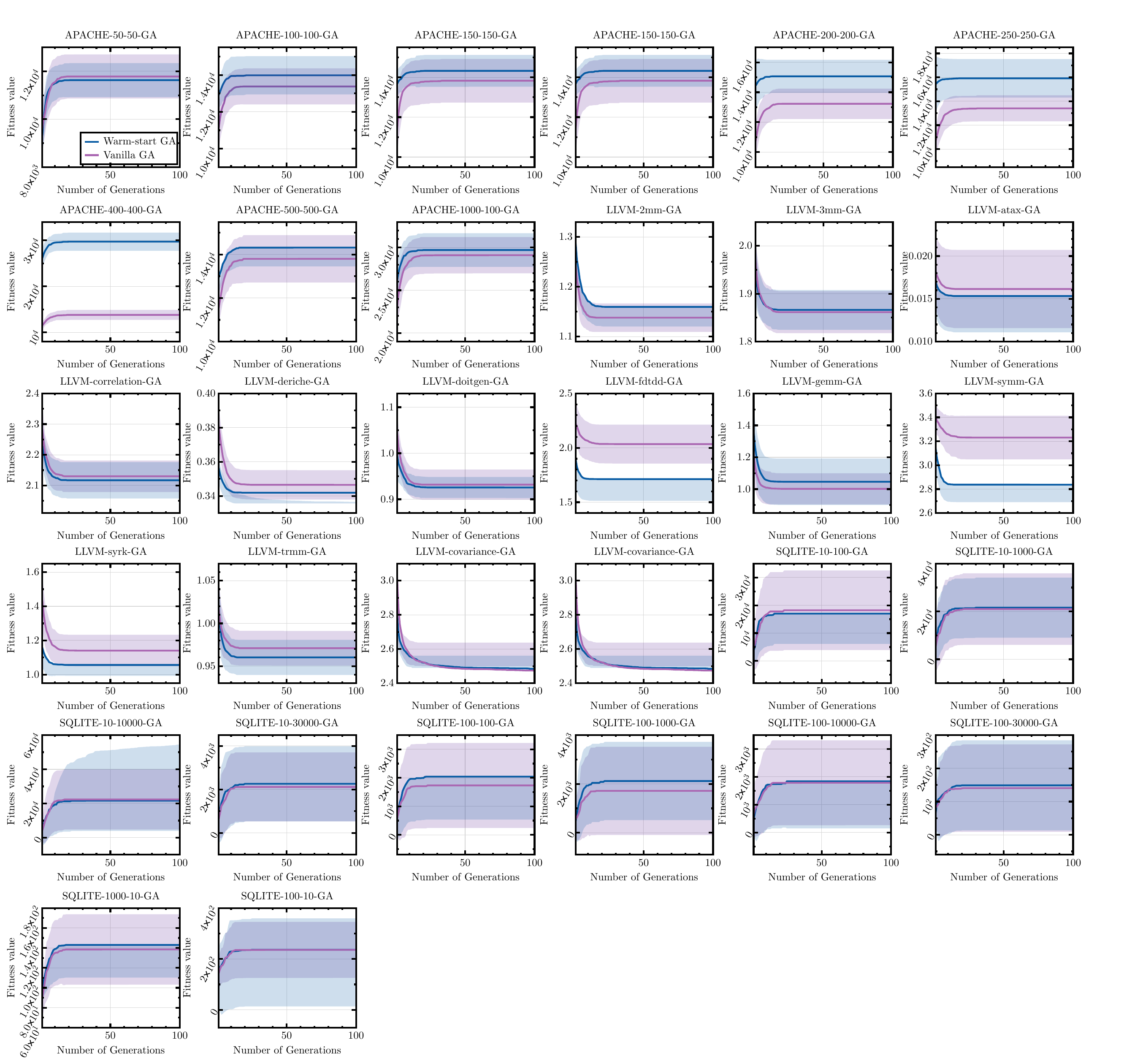}
    \caption{Evolutionary trajectories of warm-start GA (\textcolor{f3blue}{blue}) against its vanilla version (\textcolor{f3purple}{purple}) in $32$ workloads. In particular, both algorithms are started with an initialized population of $50$ and the total number of function evaluations is set to $5,000$. From these trajectories, we can see that the warm-start GA outperforms its vanilla counterpart, in terms of approximated optimal solution and the convergence rate, in over $78\%$ cases.}
    \label{fig:traj_A}
\end{figure*}

\begin{figure*}
    \centering
    \includegraphics[width=1.\linewidth]{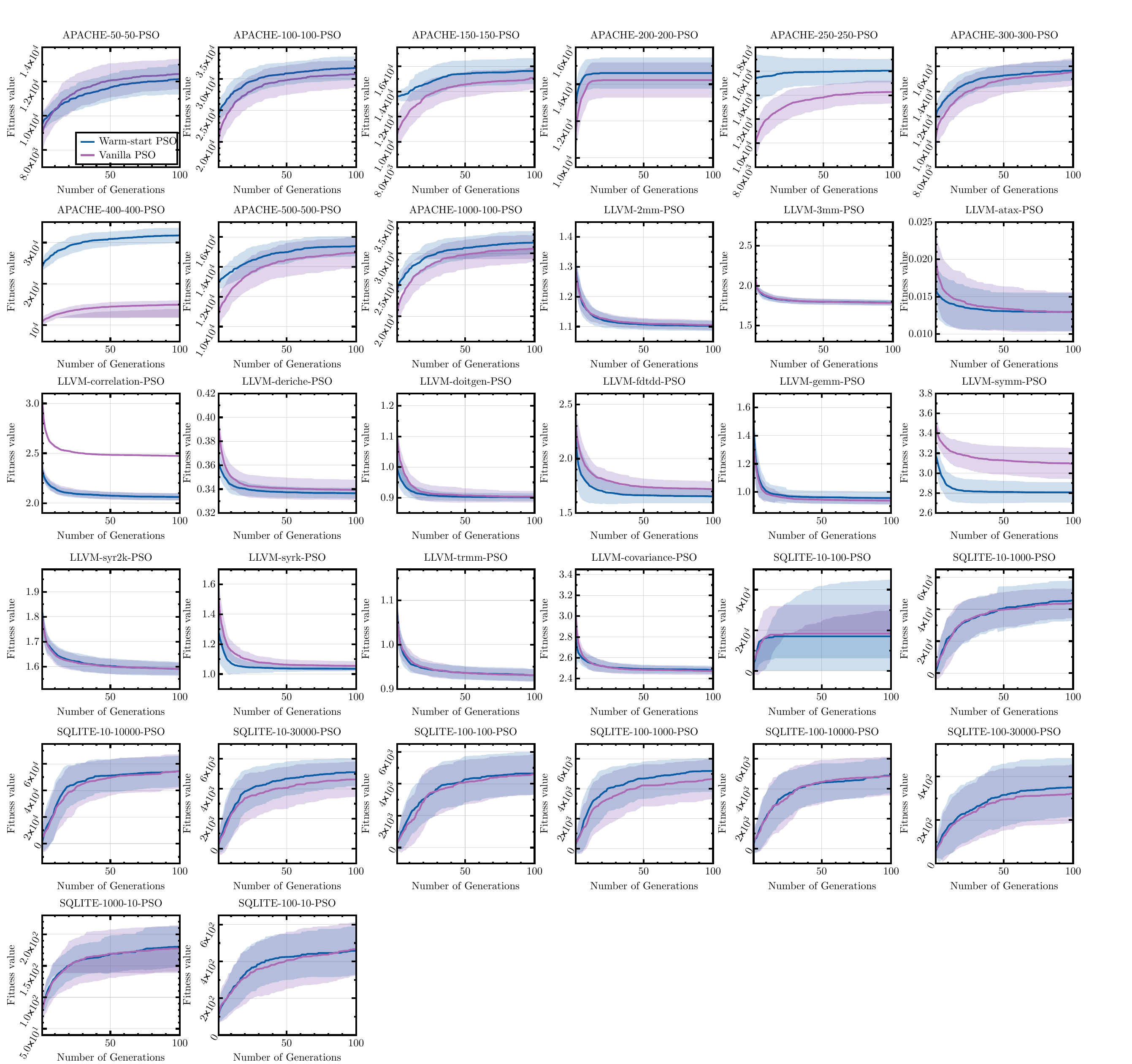}
    \caption{Evolutionary trajectories of warm-start PSO (\textcolor{f3blue}{blue}) against its vanilla version (\textcolor{f3purple}{purple}) in $32$ workloads. In particular, both algorithms are started with an initialized population of $50$ and the total number of function evaluations is set to $5,000$. From these trajectories, we can see that the warm-start PSO outperforms its vanilla counterpart, in terms of approximated optimal solution and the convergence rate, in $75\%$ cases.}.
    \label{fig:traj_B}
\end{figure*}

\begin{figure*}
    \centering
    \includegraphics[width=.4\linewidth]{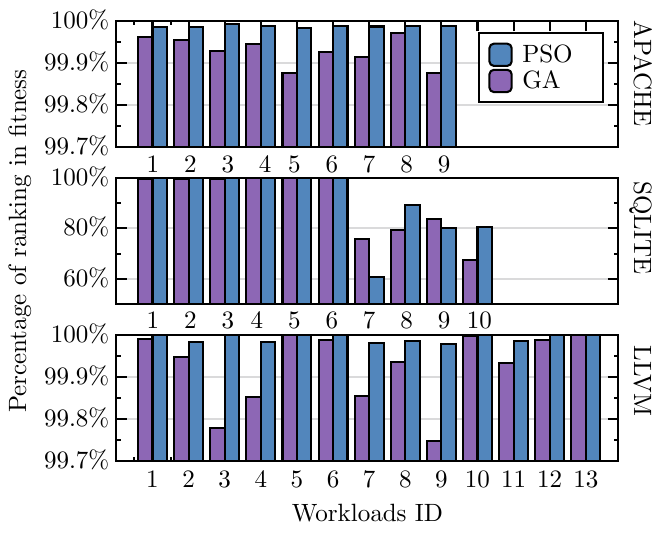}
    \caption{Bar charts of the average percentile of local optima approached by GA and PSO for \textsc{LLVM}, \textsc{Apache}, and \textsc{SQLite} on different workloads. From these comparison results, we can see that the solutions obtained by both GA and PSO are located in the top $0.1\%$ local optima whose fitness value exceeds the $99.9\%$ other local optima.}
    \label{fig:baseline}
\end{figure*}

The full results related to Sections $4.3$ to $4.4$ can be found in Figures~\ref{fig:traj_A} to~\ref{fig:baseline}.

\end{document}